\documentclass[aps,twocolumn,floatfix,showpacs,showkeys,superscriptaddress]{revtex4-1}
\usepackage{graphicx}
\usepackage{natbib}
\usepackage{epstopdf}
\usepackage{amsmath}
\usepackage{amssymb}
\usepackage{mathbbol}
\usepackage{braket}
\usepackage[hyperindex,colorlinks,urlcolor=blue]{hyperref}
\usepackage{xcolor}

\begin{document}

\title{Rare quantum metastable states in the strongly dispersive Jaynes-Cummings oscillator} 
\author{Th. K. Mavrogordatos}\email{t.mavrogordatos@ucl.ac.uk}
  \affiliation{Department of Physics and Astronomy, University College London,
Gower Street, London, WC1E 6BT, United Kingdom}
\author{F. Barratt}
 \affiliation{Department of Mathematics, Strand,
  King's College London, London, WC2R 2LS, United Kingdom}
  \author{U. Asari}
    \affiliation{Department of Physics and Astronomy, University College London,
Gower Street, London, WC1E 6BT, United Kingdom}
\author{P. Szafulski}
    \affiliation{Department of Physics and Astronomy, University College London,
Gower Street, London, WC1E 6BT, United Kingdom}
\author{E. Ginossar}
 \affiliation{Advanced Technology Institute and Department of Physics,
  University of Surrey, Guildford, GU2 7XH, United Kingdom}
  \author{M. H. Szyma\'{n}ska}
      \affiliation{Department of Physics and Astronomy, University College London,
Gower Street, London, WC1E 6BT, United Kingdom}

\begin{abstract}
  We present evidence of metastable rare quantum-fluctuation switching
  for the driven dissipative Jaynes-Cummings oscillator coupled
  to a zero-temperature bath in the strongly dispersive regime. We
  show that single-atom complex amplitude bistability is accompanied
  by the appearance of a low-amplitude long-lived transient state,
  hereinafter called the `dark state', having a distribution with {\it
  quasi}-Poissonian statistics both for the coupled qubit and cavity
  mode. We find that the dark state is linked to a spontaneous
  flipping of the qubit state, detuning the cavity to a low-photon
  response. The appearance of the dark state is correlated with the
  participation of the two metastable states in the dispersive
  bistability, as evidenced by the solution of the Master Equation and
  single quantum trajectories.
\end{abstract}

\keywords{dark state, dispersive regime, complex amplitude bistability, quantum trajectories, quantum fluctuation switching}
\pacs{42.50.Ct, 42.50.Lc, 03.65.Yz}
\date{\today}

\maketitle

\section{Introduction}
\label{sec:intro}

Fluctuation-induced switching between metastable states of driven quantum nonlinear oscillators interacting with their environment, in principle lacking detailed balance, constitutes one of the most general and intricate physics problems, intimately linked with the problem of quantum activation \cite{PeanoDykman, LeytonD, DykmanSmelyanskii, DykmanBook}. In this framework, the Duffing model provides the simplest description of a self-interacting nonlinear oscillator involving one quantum degree of freedom, where the Fokker-Planck equation (FPE) is sufficient to provide an exact treatment of quantum fluctuations. Such a formulation ultimately yields a steady state which presents notable differences from a Gaussian distribution, as one expects in linear FPEs. The FPE can be solved exactly in the steady state, since the conditions for detailed balance are satisfied at zero temperature \cite{detailbalance}. It has recently been shown that a coherently driven system with two quantum degrees of freedom, i.e., a transmon qubit coupled to a resonant cavity mode, both connected to a dissipative environment, may still be amenable to a FPE description subject to an adiabatic elimination of the fast decaying cavity field amplitude \cite{MattFPE}. 

Historically, the problem of defining switching rates between states of classical nonlinear dissipative systems is long standing. The driven Van der Pol oscillator is a very characteristic case subject to a description where an effective potential $V(x)$ can be devised as a function of the driving parameters for the nonlinear drift term of the FPE with constant diffusion. We can then define the forward (associated with an energy gap $Q_{+}$) and backward (associated with the gap $Q_{-}$) jump rates for the phase fluctuations, with a Kramers-type dependence: $r_{\pm}=[\sqrt{V^{\prime\prime}(x_0)|V^{\prime\prime}(x_b)|}/(2\pi)] \exp(-Q_{\pm}/D)$, where $x_0$ is the position of a locally-stable
potential valley, $x_b$ is the position of the barrier top, and $D$ is
the constant diffusion coefficient \cite{constantD}. 

When we analyze the single-atom Jaynes-Cummings (JC) model, we face a complexity which transcends the difficulty of solving a nonlinear FPE, as in the aforementioned oscillators. In particular, it is impossible to define an FPE to study the switching dynamics, which is a consequence of the non-perturbative nature of light-matter interaction in the strong-coupling regime \cite{CarmichaelBook1,CarmichaelBook2}. In
contrast to the dispersive optical bistability,
discussed in \cite{GrahamBist} as a special case of a dissipative
system with a potential \cite{GrahamTel}, for the case of single-atom
JC bistability we cannot formulate a suitable potential function
yielding the various attractors in the phase portrait. The inability
to obtain a potential force may lead to large deviations from the
optimal path minimizing the action for two degrees of freedom \cite{PeanoDykman, Kamenevdis}.  

\begin{figure}
\centering
\includegraphics[width=3.6in]{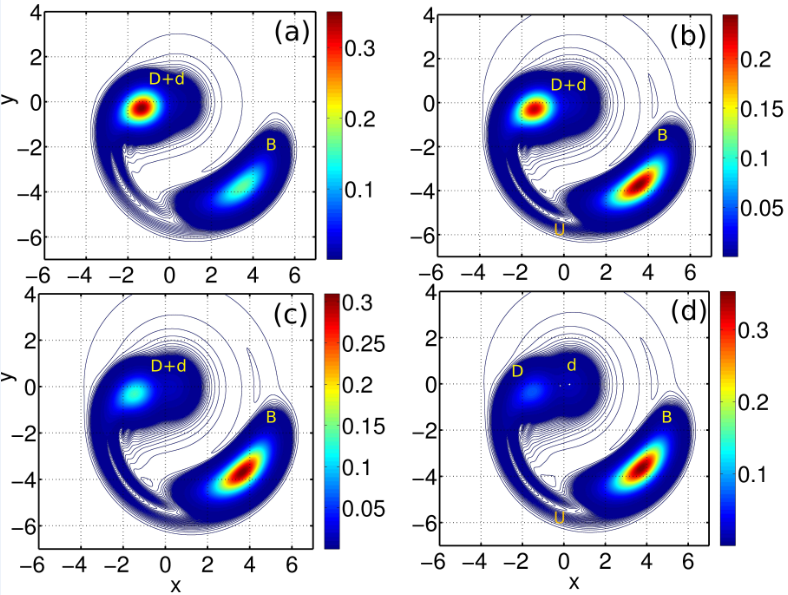}
\caption{{\bf Emergence of the dark state in the cavity field {\it quasi}-distribution.} Wigner function $W(x + iy)$ for  $\varepsilon_d/\gamma=42, 43, 44, 45$ in {\bf (a)}-{\bf (d)} respectively. $D$ denotes the dim state, $B$ the bright state, $D+d$ the complex of the coexisting dim and dark states, $d$ the dark state and $U$ the unstable mean-field state. Parameters: $\delta/g=0.873$, $\Delta\omega_c/\kappa=9.167$, $g/\gamma=600$ and $2\kappa/\gamma=12$.}
\label{fig:WignerBist} 
\end{figure}

In this paper, we report on a metastable state, called the {\it dark state}, with very low intracavity amplitude and intense qubit fluctuations which is not predicted by the mean-field equations. We identify this state as resulting from quantum bistability involving two degrees of freedom \cite{SimBistPRL}, and we depict the state on a plot of the associated {\it quasi}-distribution in the coherent phase space for the cavity field, as well as the associated distribution in the Bloch sphere (see Figs. \ref{fig:WignerBist} and \ref{fig:QT}). We find that the dark state is \textbf{(1)} rare, appearing only on the longer time scales after the transient period, \textbf{(2)} very noticeable and strongly fluctuating in the Bloch sphere with regards to the qubit observables, \textbf{(3)} long lived, compared to the typical time scales of cavity and qubit dissipation, and \textbf{(4)}
fragile to fluctuations yet more `resilient' than the unstable state
of mean-field dispersive bistability. 

In Sec. \ref{sec:methods} we define the system Hamiltonian and main methods used for analyzing quantum bistability. In Sec. \ref{sec:DOFN} we first approach with an approximate mapping to the Duffing oscillator for small drive strengths, by examining a perturbative expansion of the driven dispersive JC Hamiltonian, with renormalized parameters to account for the cavity-atom coupling. As the driving power is further increased, the Duffing approximation breaks down because both the qubit and cavity participate in the bistable switching. Subsequently, in Sec. \ref{sec:switchdyn} we study the switching dynamics in single quantum trajectories, linking our discussion to the mean-field and neoclassical predictions. While the quantum-fluctuation switching takes place in the steady state, the qubit flips and brings the cavity mode out of resonance, with a very low photon excitation. 

\section{Model and methods}
\label{sec:methods}

We will first provide a brief account of the properties of dispersive
complex amplitude bistability for varying drive strength and
frequency. In a frame rotating with the drive frequency $\omega_d$,
the Hamiltonian describing the interaction of a damped two-level atom
(qubit), with inversion operator $\sigma_z$ and raising (lowering)
operators $\sigma_{+}$ ($\sigma_{-}$) with bare resonant frequency
$\omega_q$, and a driven damped cavity mode (with photon annihilation
and creation operators $a$ and $a^{\dagger}$ respectively) with bare
frequency $\omega_c$, reads \cite{WallsBook}:
\begin{equation}\label{JC}
\begin{aligned}
H_{\rm{JC}}&=-\hbar\Delta \omega_c a^{\dagger}a - \frac{1}{2} \hbar\Delta \omega_q \sigma_z\\
&+ i\hbar g(a^\dagger \sigma_{-} - a\sigma_{+}) + i\hbar( \varepsilon_d a^{\dagger} -\varepsilon_d^* a),
\end{aligned}
\end{equation}
where $\Delta\omega_{c,q}=\omega_d-\omega_{c,q}$, $g$ is the
atom-cavity coupling strength and $\varepsilon_d$ is the drive
amplitude (or strength). The cavity is
coupled to a thermal bath at zero temperature inducing a photon loss
rate of $2\kappa$, while the qubit relaxation rate is denoted by $\gamma$ (due to both radiative and non-radiative processes, such as quasi-particle formation). 

The system density matrix obeys the Lindblad master equation (ME) \cite{WallsBook}:
\begin{equation}\label{ME}
\begin{aligned}
&\dot{\rho}=[1/(i\hbar)][H_{\rm{JC}}, \rho] + \kappa \left(2a\rho a^{\dagger}-a^{\dagger}a \rho - \rho a^{\dagger}a\right) \\
&+ (\gamma/2) \left(2\sigma_{-}\rho \sigma_{+} - \sigma_{+}\sigma_{-}\rho - \rho \sigma_{+} \sigma_{-}\right),
\end{aligned}
\end{equation}
which is solved numerically via exact diagonalization, as well as unraveled into single quantum trajectories. The steady-state solution of \eqref{ME} $\rho_{\rm ss}$ is used to calculate the average value of the observables considered, as $\braket{O}_{\rm ss}={\rm tr}(\rho_{\rm ss}O)$, where $O$ is a system operator. A normalized conditional state unraveling the ME is subject to an evolution obeying the Stochastic Schr\"{o}dinger Equation (SSE):
\begin{equation}\label{SSE}
d\psi_{k}(t)=D_1[\psi_{k}(t)]dt + D_2[\psi_{k}(t)]dW(t),
\end{equation}
where $D_1$ is the drift term, $D_2$ is the diffusion term (both functions of Lindblad operators), and $dW$ is a real increment (for more details, see \cite{PlatenBook, OpenQ}). The density matrix $\rho_{k}(t)=\ket{\psi_{k}(t)}\bra{\psi_{k}(t)}$ is used to calculate the average value of system observables as $\braket{O(t)}={\rm tr}(\rho_{k}(t)O)$ \cite{CarmichaelBook2, PlatenBook}. In all of the cases considered here, the initial state $\psi_{k}(t=0)$ is a pure state with $\braket{\psi_{k}(0)|a^{\dagger}a|\psi_{k}(0)}=0$ and $\braket{\psi_{k}(0)|\sigma_{z}|\psi_{k}(0)}=-1$, unless explicitly stated otherwise. The properties of steady-state bistability were not affected by a change in the initial conditions {\it for single quantum trajectories}. Convergence with respect to the time step in the evolution as well as in the number of states comprising the truncated Hilbert space for the cavity has been ensured.

The {\it strongly dispersive regime} is defined by an atom-cavity detuning $\delta
\equiv |\omega_c -\omega_q|$ of the order of (and usually greater than) the coupling strength $g \gg 2\kappa, \gamma$, alongside its relation to the coherent drive strength: ${\rm max}(2\kappa, \gamma)<\varepsilon_d \ll g^2/\delta$, which takes us beyond the linear dispersive regime. For the detuning that we are considering here, $\Delta \omega_q>g$ and $2\kappa<\Delta \omega_c<g^2/\delta$. For all cases discussed in this work, $\delta \ll \omega_c + \omega_q$ while $\omega_c, \omega_q \gg g$, so that the rotating wave approximation (RWA) can be safely performed (see also Fig. 1 of \cite{RWAval}) for the number of photons involved in the steady-state response. For the average photon number $\braket{n}_{\rm \small ss}$ in the steady state, we would typically have $g \sqrt{\braket{n}_{\rm \small ss}+1} \leq 0.1\, \omega_c$. The standard ME can then adequately describe many cavity QED and circuit QED experiments \cite{RedME}, both at resonance and in the dispersive regime (for a direct comparison between theory and experiment see e.g. \cite{BishopNL} and \cite{SimBistPRL}). 

In our treatment we have not included the phase destroying term $(\gamma_{\phi}/2)(\sigma_{z}\rho \sigma_{z}-\rho)$ in the ME \cite{CarmichaelBook1}. Such a term produces the decay coefficient $\gamma+2\gamma_{\phi}$ for the qubit coherence, erasing even more rapidly the memory of the initial state in the averaged system response. Qubit dephasing would affect the lifetime and fluctuations of the states of quantum bistability together with the scaling constants of the mean-field equations, both present when the (energy) decay coefficient $\gamma$ is already taken into account. For the limiting cases considered later on (see Fig. \ref{fig:FFTs}), we would take $\gamma \to 0$ together with $\gamma_{\phi} \to 0$.

\section{From the effective Duffing oscillator to the full JC nonlinearity}
\label{sec:DOFN}

\subsection{The effective Duffing oscillator}
\label{subsec:effectiveDuff}

The quantum Duffing oscillator is a precursor of the JC nonlinearity. After applying
the dispersive transformation \cite{DispersiveTransform, CanonicalTr},
the Hamiltonian of Eq. \eqref{JC} in the dressed-cavity Duffing
approximation for $\delta \gg g$ reads (up to quartic order in the small parameter
$g/\delta$):
\begin{equation}\label{Duffingapprox}
\begin{aligned}
H_{\rm D}& = \hbar \left(\omega_c -\omega_d + \frac{g^4}{\delta^3} -
  \frac{g^2}{\delta}\sigma_z + 2 \frac{g^4}{\delta^3}\sigma_z
\right)a^{\dagger}a \\
&+\hbar \frac{g^4}{\delta^3}\sigma_z{a^{\dagger}}^2 a^2 + i\hbar(\varepsilon_d a^{\dagger} -\varepsilon_d^* a).
\end{aligned}
\end{equation}
\begin{figure*}
\centering
\includegraphics[width=6in]{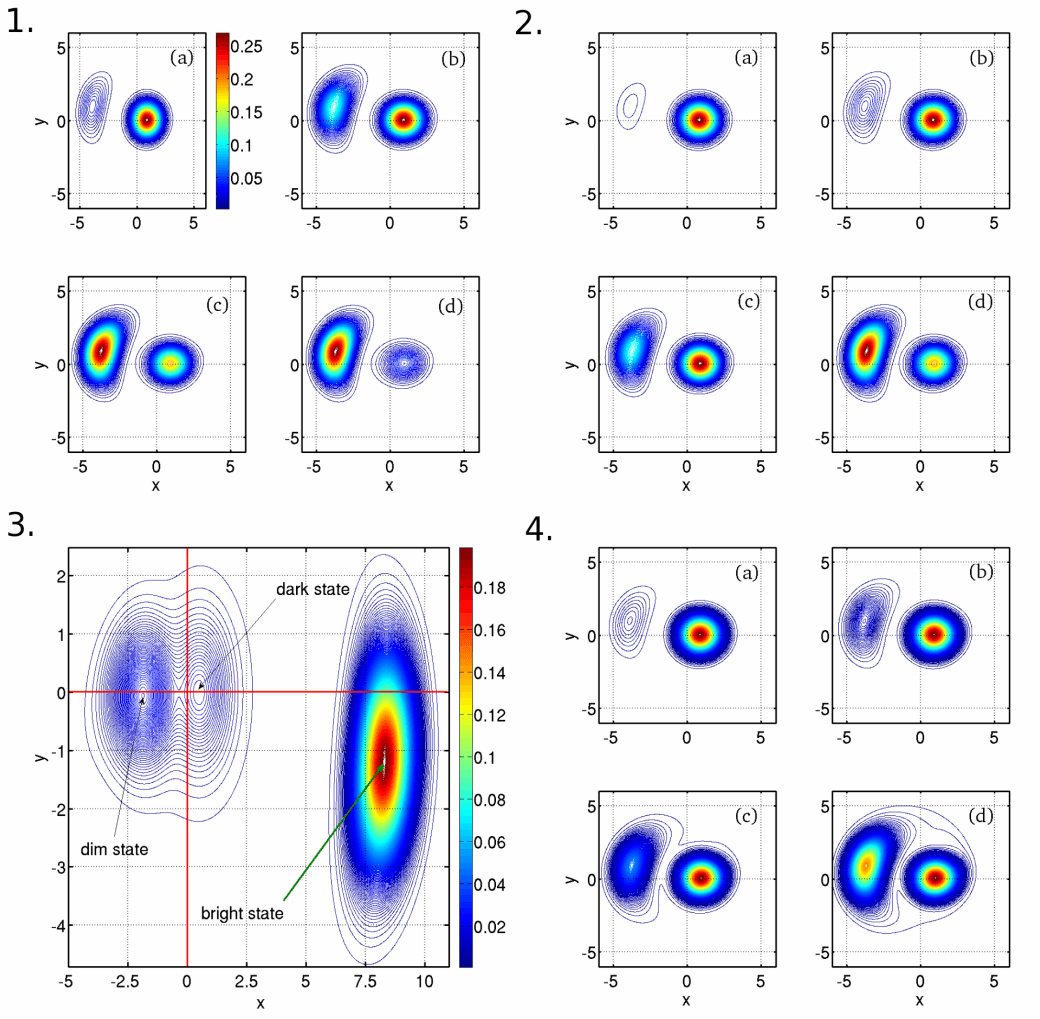}
\caption{\textbf{Cavity bimodality and the dark state.} ME results for the \textit{quasi}-distribution function $Q(x + iy)$ for four equispaced values of the driving frequency in the interval $\Delta\omega_d/\kappa=[55.83, 57.50]$ corresponding to the frames (a)-(d) for \textbf{Panels 1 and 2}. In \textbf{Panel 1}: $\varepsilon_d/\gamma=100$, $2\kappa/\gamma=12$ and in \textbf{Panel 2}: $\varepsilon_d/\gamma=95$, $2\kappa/\gamma=12$. In \textbf{Panel 3}: $Q$ function $Q(x + iy)$ for $\varepsilon_d/\gamma=350$, $2\kappa/\gamma=12$, $\Delta\omega_c/\kappa=37.50$ and in \textbf{Panel 4}: \textit{quasi}-distribution $Q$ function $Q(x + iy)$ for the same values of the driving frequency and the same parameters as for Panel 1, but with $2\kappa/\gamma=0.25$.}
\label{fig:Q}
\end{figure*}  
In the above expression we have kept only linear terms with respect to
$g/\delta$ in the transformed drive term, while we have set
$\sigma_{\pm}=\braket{\sigma_{\pm}}=0$, taking $\sigma_z=\braket{\sigma_z}\approx -1$ (see Eq. 3.15 of \cite{DispersiveTransform}). Based on Eq. \eqref{Duffingapprox}, we
can extract the Wigner function for the effective Duffing oscillator
\cite{WignerKheruntsyan, WignerKheruntsyanOneOsc}, calculated via the
generalized $P$-representation (see the Appendix for a
full derivation):
\begin{equation}\label{WignerFinal}
  W(\alpha, \alpha^{*})=\frac{2}{\pi} e^{-2|\alpha|^2} \frac{\left| _0F_1 \left(c, 2 \tilde{\varepsilon}_d \alpha^{*} \right) \right|^2}{_0F_2(c,c^*,2| \tilde{\varepsilon}_d|^2)},
\end{equation}
where $_0F_1(a;x)$ and $_0F_2(a,b;x)$ are generalized hypergeometric
functions of the variable $x$ with parameters $a, b$. Here, $c=(\kappa
- i\Delta \omega_c^{\prime}) /(i\chi)$ with the renormalized detuning
$\Delta \omega_c^{\prime}=\Delta\omega_c+(g^2/\delta)\sigma_z -
(g^4/\delta^3)(2\sigma_z+1)$ and $\chi=(g^4/\delta^3) \sigma_z$, while
$\tilde{\varepsilon}_d=\varepsilon_d/(i\chi)$. The Wigner distribution
function of Eq. \eqref{WignerFinal} is valid for $4\braket{N}_{\rm ss}g^2/\delta^2 \ll 1$ (here the subscript ${\rm ss}$ denotes the steady state, and $N=a^{\dagger}a + \sigma_{+}\sigma_{-}$ is the number operator of
system excitations; see \cite{DispersiveTransform} and \cite{BishopJC} for more details) and can be used for the calculation of the
intracavity field moments in the complex plane as opposed to the
four-dimensional space of \cite{DuffingWalls}.  Valuable information can be extracted from the effective Duffing oscillator model for low drive strengths, where $\sigma_z \approx \braket{\sigma_z} \approx -1$. The expression of Eq. \eqref{WignerFinal} for the Wigner function of the renormalized Duffing oscillator predicts a variety of critical points surrounding the dim state (in agreement with the low-amplitude bistability plots presented in \cite{WignerKheruntsyanOneOsc}). The steady-solution of the ME yields an almost identical distribution, capturing the same amount of nodes in very similar positions. In the regime of low intracavity amplitude, quantum fluctuations are essential for the onset of complex amplitude bistability, determined by the scale parameter $\delta^2/(4g^2)$. With increasing drive strength, where the number of system excitations approaches the scale parameter, the Duffing approximation becomes inapplicable, and the full JC dynamics with two quantum degrees of freedom must be taken into account.
\begin{figure}
\centering
\includegraphics[width=3.6in]{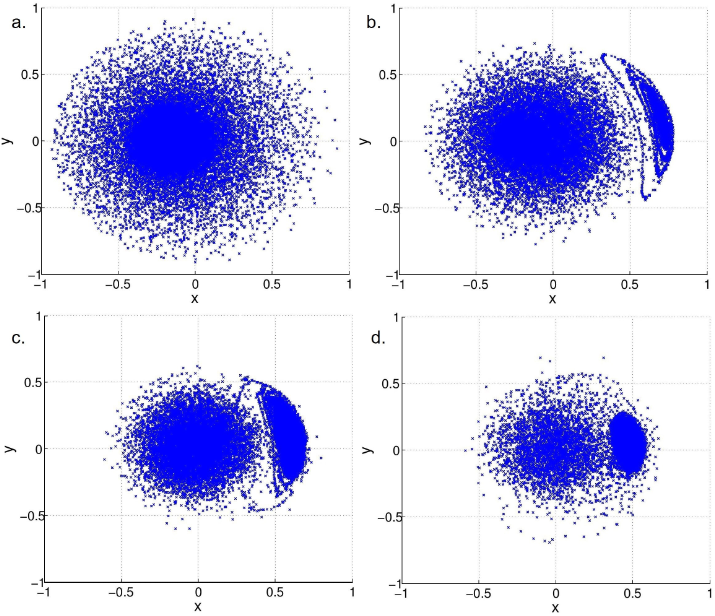}
\caption{Projections on the spin-x and spin-y axes of the Bloch sphere equatorial plane from single quantum trajectories and varying drive frequency (corresponding to a horizontal cut in the semiclassical bistability leaf pictured in Fig. \ref{fig:3cuts}). Parameters : $g/\delta=0.14$,  $(2 \kappa)/\gamma = 12$, $g/\gamma = 3347$, ${\varepsilon_d}/{(\gamma)} = 100$ with \textbf{(a)} $ \Delta\omega_c/\kappa =47.500$, \textbf{(b)} $ \Delta\omega_c/\kappa=56.667$, \textbf{(c)} $\Delta\omega_c/\kappa=65.833$, and \textbf{(d)} $\Delta\omega_c/\kappa=75.000$.}
\label{fig:SupF2B} 
\end{figure}
For the driven JC oscillator, the semiclassical bistability region,
characterized by one unstable and two metastable states (a {\it dim} state with lower photon occupation and a {\it bright} state
with a higher photon number), is constructed in the drive parameter
space $(\Delta\omega_c/\kappa, \varepsilon_d/\kappa)$ from the
Maxwell-Bloch equations \cite{WallsBook}. The latter are known to
yield solutions that exhibit overlap between different domains of
attraction and chaotic behavior \cite{BifRoutes, AttractorsLaser}. An
alternative construction can be carried out from Hamilton's equations
of motion for time scales during which the qubit degrees of freedom
can be considered as constants of motion for $\gamma/(2\kappa) \to 0$
\cite{BishopJC}.
\begin{figure*}
\centering
\includegraphics[width=6.0in]{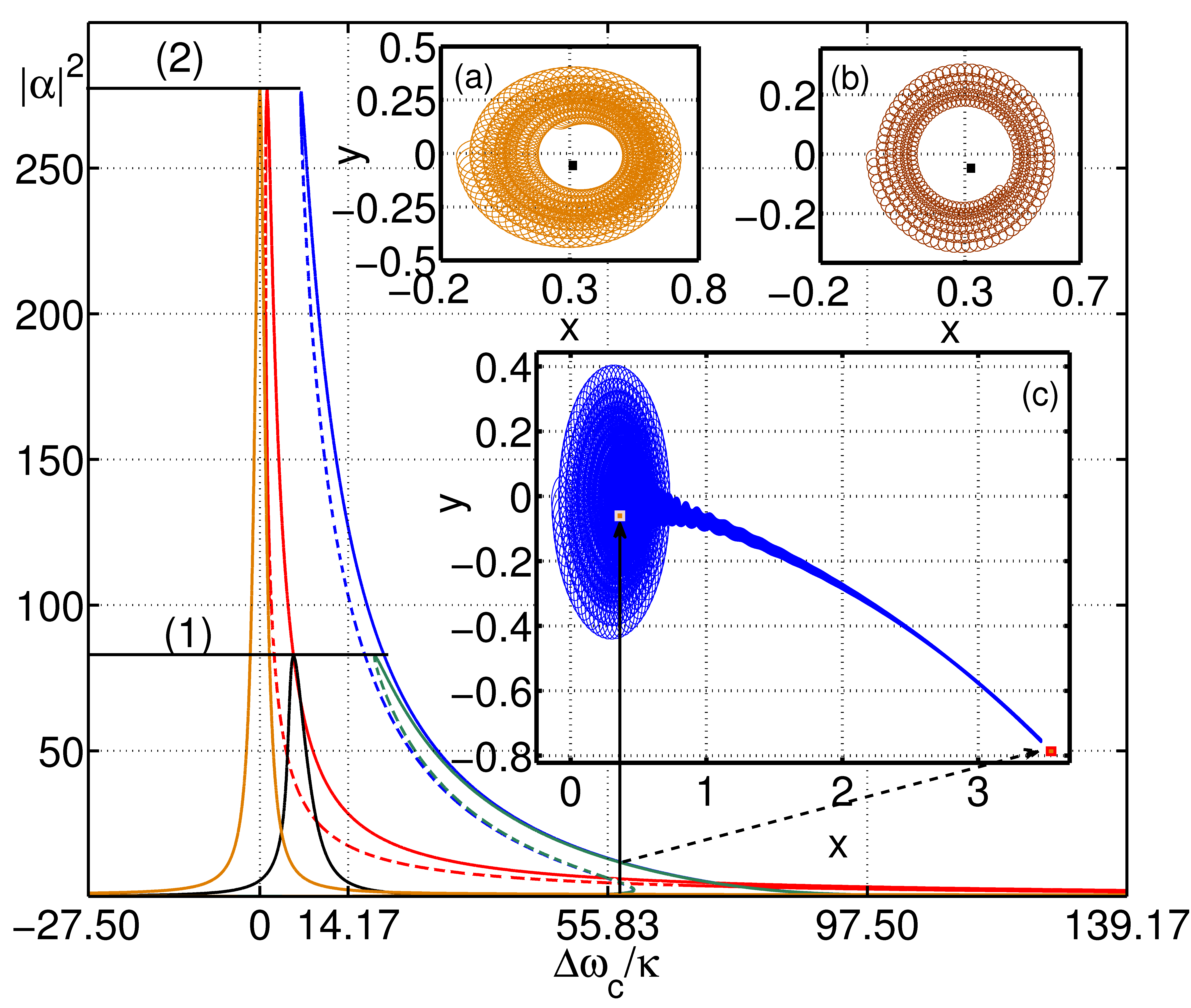}
\caption{\textbf{Steady-state solution of the Maxwell-Bloch equations for the intracavity amplitude in the presence of spontaneous emission.} The solid curves depict Lorentzian and skewed-Lorentzian profiles for the same drive amplitude, $\varepsilon_d/\gamma=100$: \textbf{\textit{blue, skewed Lorentzian, with peak at level (2) further from resonance ($\Delta \omega_c=0$)}} for $g/\delta=0.14$ and $2\kappa/\gamma=12$; \textbf{\textit{red, skewed Lorentzian, with peak at level (2) closer to resonance}} for $g/\delta=0.87$ and $2\kappa/\gamma=12$; \textbf{\textit{green, skewed Lorentzian, with peak at level (1)}} for $g/\delta=0.14$ and $2\kappa/\gamma=22$; \textbf{\textit{black, Lorentzian with peak at level (1)}} for $g/\delta=0.042$ and $2\kappa/\gamma=22$; \textbf{\textit{orange, Lorentzian with peak at level (2)}} for $g/\gamma=3347$, $\delta/g=0$ and $2\kappa/\gamma=12$. The ratio $\varepsilon_d/\kappa$, giving the empty cavity amplitude, determines the plateau for the Lorentzian peaks marked by (1) and (2) respectively. The peak of the orange curve indicates the bare cavity frequency. The three insets (a, b, c) represent solutions of the Maxwell-Bloch equations of the phase space $x-y$ of the intracavity field for varying time, corresponding to the drive frequency $\Delta\omega_c/\kappa=56.83$. In \textbf{(a)} $g/\gamma=3347$ and in \textbf{(b)} $g/\gamma=1000$. In \textbf{(c)} we are plotting the intracavity field for $g/\gamma=3347$ but for a longer time, showing the approach of the second semiclassical state (bright) in a limit cycle fashion. The dashed lines indicate the unstable branches.}
\label{fig:SCph}
\end{figure*}
Taking now into account the quantum fluctuations,
Fig. \ref{fig:WignerBist} shows the onset of complex amplitude
bistability extracted from the ME solution for the
steady-state intracavity field for a constant drive detuning and
varying strength. The two metastable states, the dim (D) and the
bright (B), exchange probability as the drive strength increases,
being connected via the dispersive excitation spiral along which we
can also discern the unstable state (U). Together with the increase of
the intracavity photons, we can also observe the progressive
separation between the dim and the dark states, which will also be
demonstrated when the ME is unraveled into single trajectories. A
good separation between the dim and the dark states is shown in Fig. \ref{fig:WignerBist}(d). In the following, we will also discuss the behavior of the qubit observables when the perturbative approach leading to the Hamiltonian of Eq. \eqref{Duffingapprox} cannot be applied.
\begin{figure}
\centering
\includegraphics[width=3.5in]{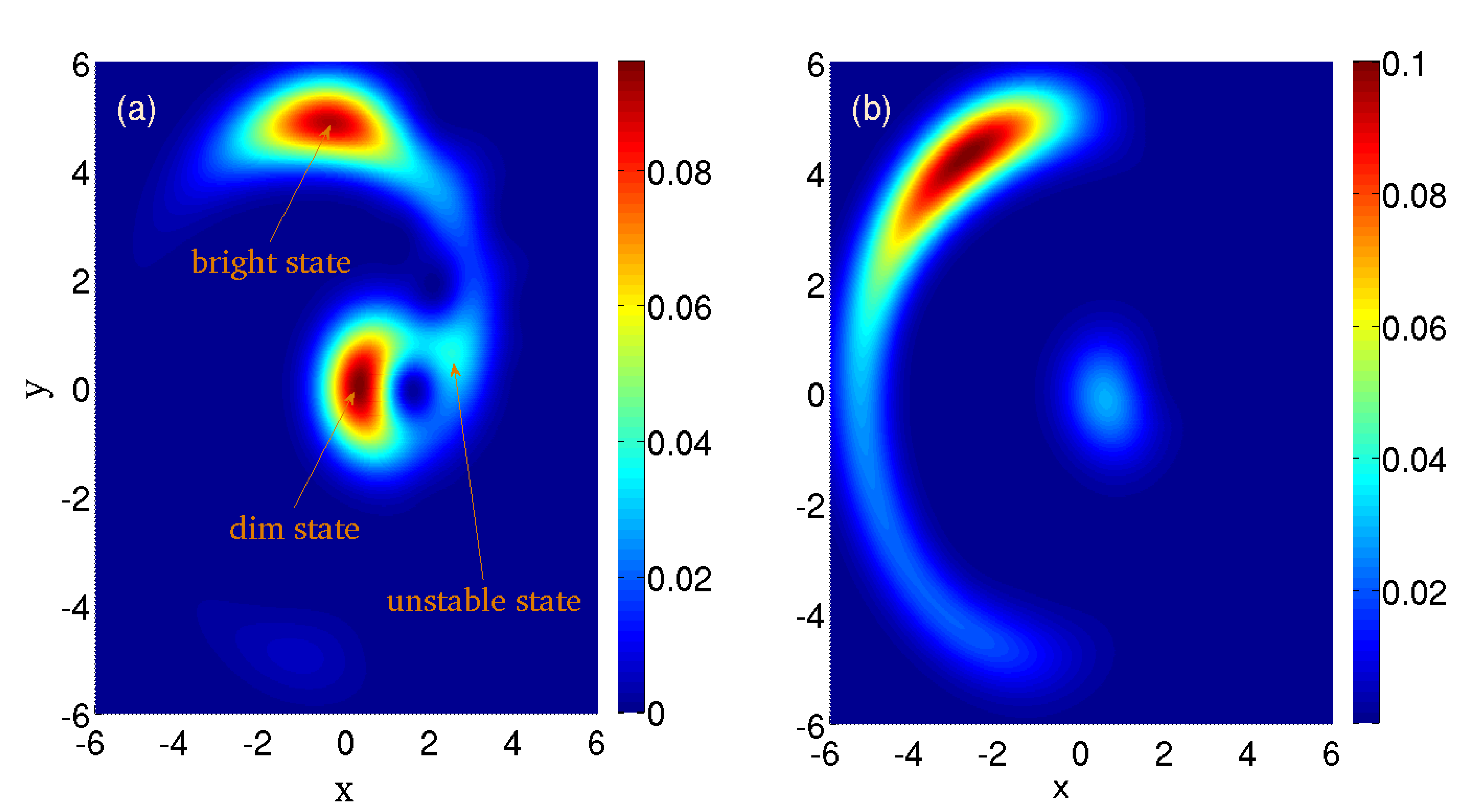}
\caption{\textbf{The dispersive cavity excitation spiral}. \textit{Quasi}distribution function $Q(x + iy)$ for two time instants $t_1$ \textbf{(a)} and $t_2$ \textbf{(b)} with $t_2>t_1$ during a switch to the bright metastable state. Parameters: $\varepsilon_d/\kappa=16.67$, $g/\delta=0.14$, $\gamma/(2\kappa)=1/12$, $g/\gamma=3347$.}
\label{fig:Qfinal} 
\end{figure}

\subsection{Mean-field and quantum trajectories away from the critical point $C_1$}
\label{subsec:meanfC1}

In order to understand the origin of the dark state we have at first ignored the cavity-qubit quantum correlations. While assessing single-atom dispersive bistability, the authors in
\cite{BishopJC} present a construction in which they depict the
semiclassical bistability region as a `leaf' in the phase space when
$\gamma \to 0$, opening at one critical point ($C_1$) in agreement with
the effective Duffing oscillator, and closing at another ($C_2$) when
the drive is at resonance with the significantly excited
cavity. Quantum fluctuations of both the qubit and cavity field alter
the overall shape of the leaf, bringing about non-equilibrium dynamics
where quantum noise cannot be treated perturbatively. As the drive strength and the intracavity photon number are further increased, the {\it quasi}-distribution function of Eq. \eqref{WignerFinal} fails to adequately describe the quantum dynamics, which now involve the qubit more actively. The region of coexisting states with probabilities of the same order of magnitude marks the boundary of the region where quantum fluctuations are important. Solving the ME for the system density matrix in the steady state, and tracing out the qubit degrees of freedom, we can investigate how the cavity bistability builds up with varying drive frequency and power. The mean-field predictions and the full quantum treatment are in closer agreement outside the bistability region. Inside this region, conversely, we expect a first-order quantum phase transition boundary marked by coexistent semi-coherent states, clearly indicated by the $Q$ function plots we present in Fig. \ref{fig:Q}. One metastable state has a low photon mean $\braket{a^{\dagger}a}$, called `dim' while the other one has a higher cavity excitation, called `bright'. 
\begin{figure}
\centering
\includegraphics[width=3.7in]{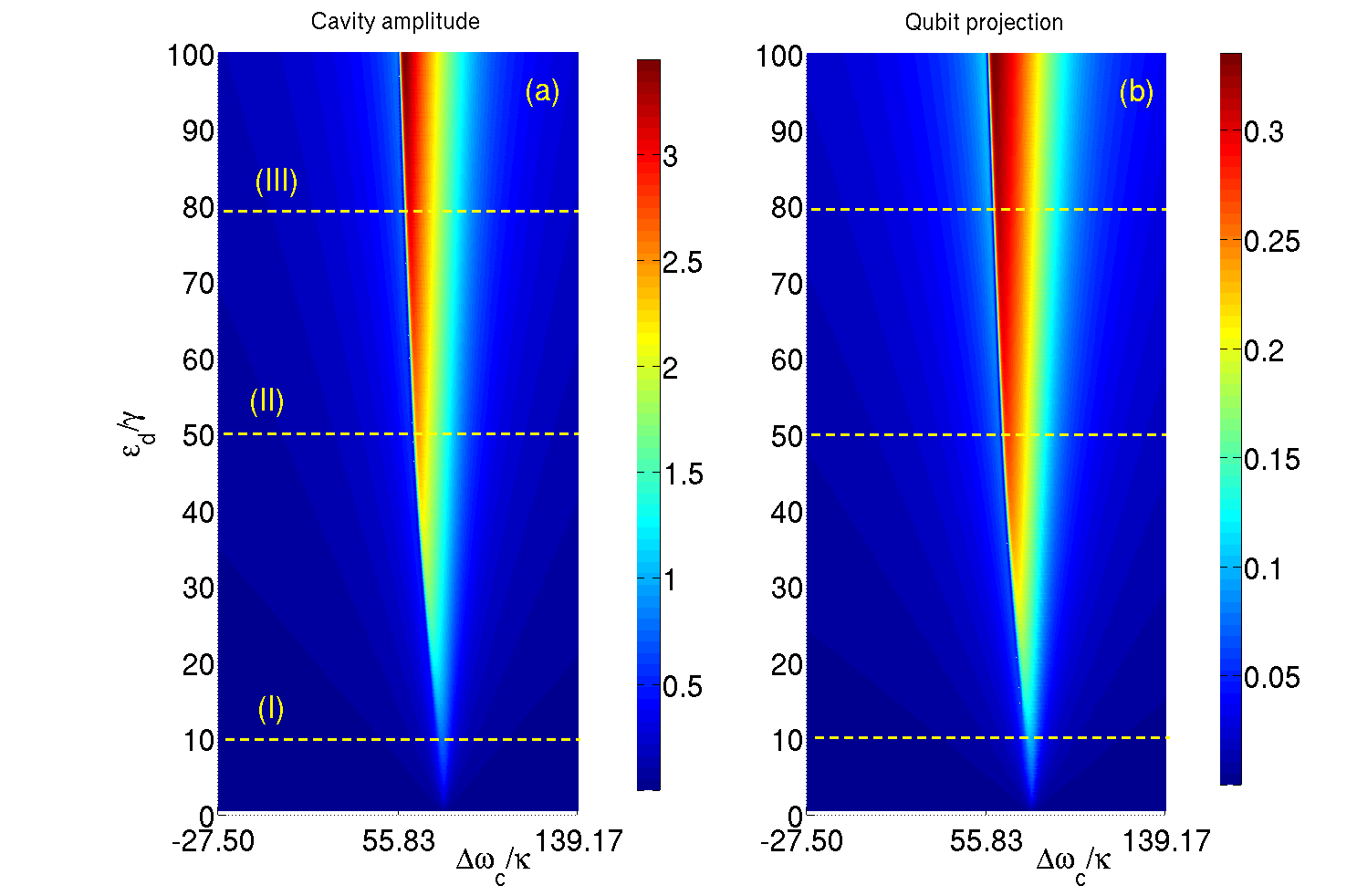}
\includegraphics[width=3.0in]{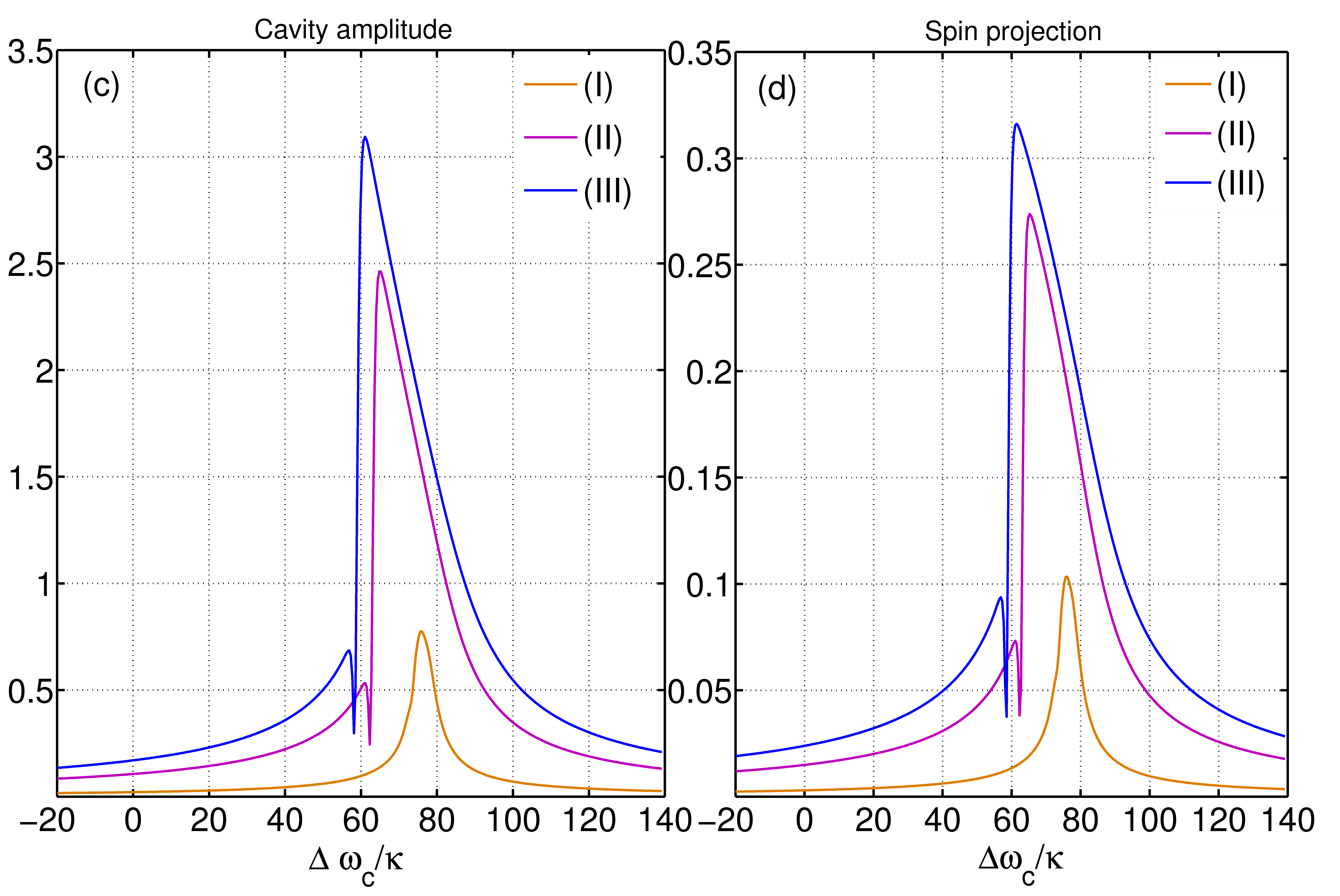}
\caption{\textbf{Coherent cancellation in the JC oscillator response.} Intracavity photon field $\left|\braket{a}\right|$ in \textbf{(a)} and qubit projection $\left|\braket{\sigma_{-}}\right|$ in \textbf{(b)} for the steady state solution of the JC model for varying drive frequency and strength. The development of the coherent cancellation with increasing drive strength for the three cuts, I, II and III, is depicted in {\bf (c)} and {\bf (d)} for the cavity field and qubit, respectively. Parameters: $g/\delta=0.14$, $\gamma/(2\kappa)=1/12$, $g/\gamma=3347$.}
\label{fig:ME2f} 
\end{figure}
In Panels 1 and 2 of Fig. \ref{fig:Q} we are traversing the steady-state quantum bistability region by varying the drive frequency at constant drive strength. The dim coherent state gives its place to the bright one while crossing the first-order transition line. In Panel 3 we present evidence of the dark state in the averaged response, where we are able to discern a center and a saddle point. The state appears to be adjacent and linked to the dim metastable state. In Panel 4 we attempt a pictorial analogy to the case where the photon loss rate $2\kappa$ is of the same order of magnitude as the spontaneous emission rate $\gamma$ [going further away from the {\it zero system size}, i.e. the limit $\gamma^2/(8g^2)=0$], revealing that the bright and dim states are joined in probability transfer as a consequence of increased spontaneous emission. Furthermore, the variation of $\gamma$ has an important effect on the steady-state distribution, resulting in the persistence of the dim state for the same drive strength (compare frames 1 and 4), as opposed to the low-power bistability. In that respect, a high photon number limiting behavior for this system far from equilibrium can be defined through the intracavity amplitude $n_{\rm scale}=[\delta/(2g)]^2$ for which the nonlinearity in the response can no longer be treated perturbatively \cite{PhotonBlockade}. In the strongly dispersive regime, the presence of the small term $g/\delta$ precludes the divergence of nonlinearity at low intracavity amplitudes in the \textit{a priori} absence of spontaneous emission and dephasing ($\gamma, \gamma_{\phi}$=0), as deduced from \cite{PhotonBlockade}.
\begin{figure}
\centering
\includegraphics[width=3.7in]{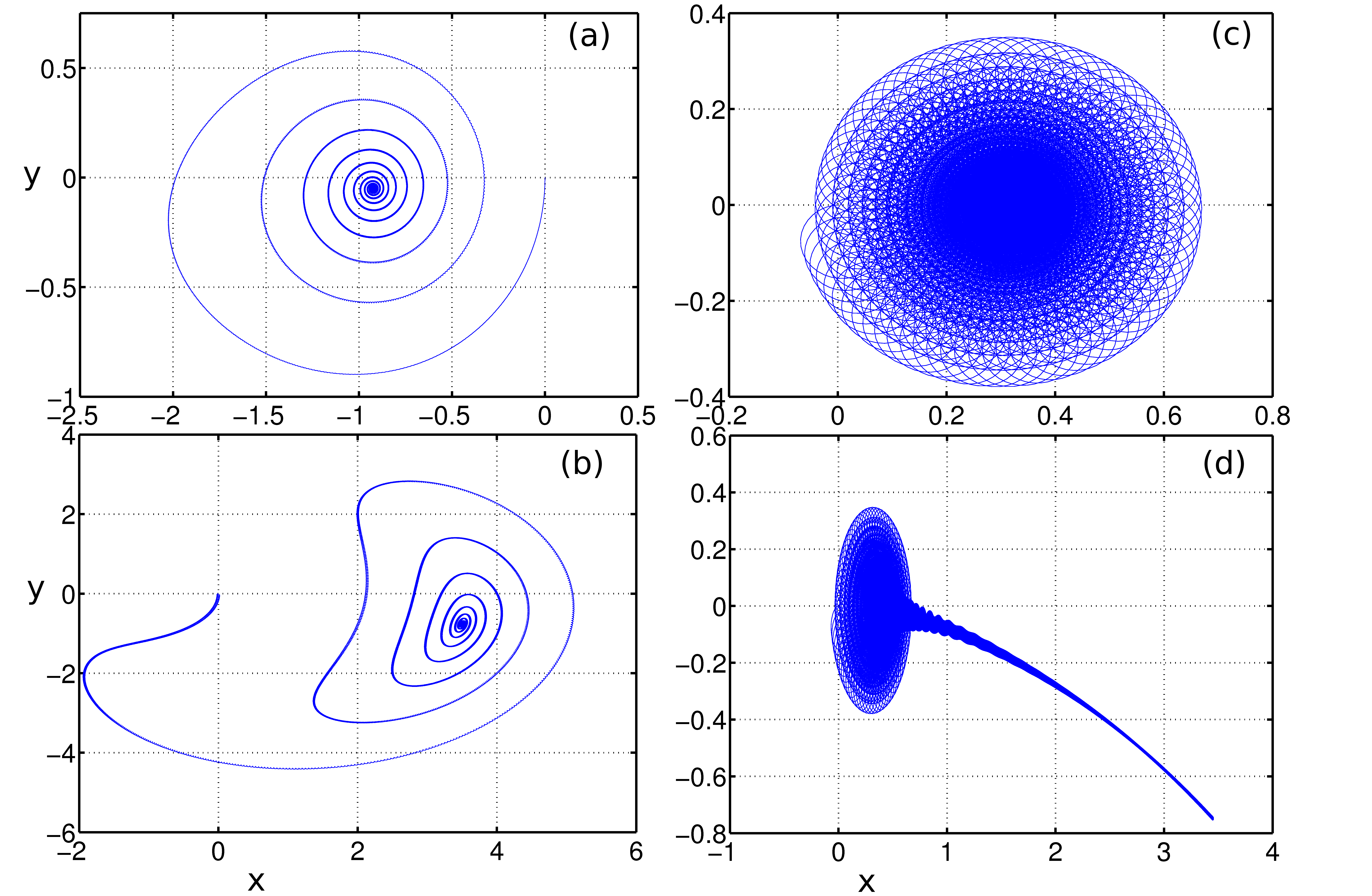}
\caption{Intracavity mean-field phase portraits $(x(t), y(t))$, including the transient response, for different initial values of the qubit Bloch vector $\mathbf{s}(0)=(\braket{\sigma_{x}(0)}, \braket{\sigma_{y}(0)},\braket{\sigma_{z}(0)})$. \textbf{(a)} $\gamma=0$, $\gamma < 2\kappa \neq 0$, $\mathbf{s}(t=0)=(0, 0, -1)$. \textbf{(b)} $\gamma=0$, $\mathbf{s}(0)=(0, -\sqrt{1-0.95^2}, -0.95)$.  \textbf{(c)} $\gamma=0$, $\mathbf{s}(t=0)=(0, -1, 0)$. \textbf{(d)} $\gamma < 2\kappa \neq 0$, $\mathbf{s}(t=0)=(0, -1, 0)$. In all cases $(x, y)(t=0)=(0, 0)$.  Parameters: $\Delta\omega_c/\kappa=56.833$, $\varepsilon_d/(2\kappa)=100/12$, $2\kappa/\gamma=12$.}
\label{fig:SupF3} 
\end{figure}
We will now focus on the switching behavior of a qubit coupled to a driven resonant cavity mode in the regime of dispersive bistability far away from the critical point $C_1$ in the semiclassical bistability leaf. Qubit switching is revealed by single quantum trajectories after calculating the reduced qubit density matrix, $\rho_{k\,Q}(t)={\rm tr}_c \{\ket{\psi_{k}(t)} \bra{\psi_{k}(t)}\}$. The dynamical organization of the {\it quasi}-coherent distributions is depicted in the Bloch sphere, which serves as an equivalent representation of the phase space for the complex cavity amplitude. Figure \ref{fig:SupF2B} shows the build-up of bistability in the qubit amplitude projected on the equatorial plane for constant driving power and varying driving frequency. Each point in the scatter plots corresponds to one time instant within a single quantum trajectory, generated by solving numerically an SSE under the diffusive approximation using an explicit weak scheme \cite{OpenQ, PlatenBook}. 

This representation is analogous to the development of cavity bimodality depicted in the Wigner function plots presented in Fig. \ref{fig:WignerBist}. The dim state exhibits a concentration around the south pole of the Bloch sphere while the bright state approaches the equatorial plane with decreasing drive-cavity detuning $\Delta \omega_c$. As $\Delta \omega_c$ increases, the bright state distribution moves towards the south pole, consistent with the approach of the Lorentzian lineshape outside the bistability leaf, where only one state (and consequently one distribution) is expected. An example of the mean-field distorted Lorentzian profiles within the bistability region are given in Fig. \ref{fig:SCph}. 

In Fig. \ref{fig:SCph} we show the mean-field dynamics for $\gamma \neq 0$ alongside the nonlinear cavity response in the steady state (see Ch. 11 of \cite{WallsBook} for the relevant equations). Here we can observe the decreasing nonlinearity for decreasing coupling strength, marking the transition from a skewed Lorentzian curve to a linear response function. A solid line with an arrowhead intersects the blue curve at three points: the dim, the unstable and the bright states from bottom to top respectively. Mean-field (Maxwell-Bloch) bistability is present in the transient evolution as well, evidenced by the two distinguished modes in the Inset (c), exhibiting anew the limit cycle approach we encounter for the qubit in one single quantum trajectory. The Insets (a) and (c) focus on the approach of the dim state (marked with a point), with varying coupling strength.

Focusing now on a quantum trajectory, in Fig. \ref{fig:Qfinal} we present a field \textit{quasi}-distribution function for the reduced cavity density matrix $\rho_{\rm k\,C}(t)={\rm tr}_{\rm Q}\{\ket{\psi_{k}(t)} \bra{\psi_{k}(t)}\}$ at two time instants $t_1, t_2$ during a period of macroscopic switching of the coupled cavity-qubit system to the bright state. We find that the three semiclassical states coexist along a spiral during a switch `up' to the bright state. This figure, moreover, shows that the dim and the unstable (semiclassical) state are connected by two probability-flow paths, similar to the development of bistability we have seen in Fig. \ref{fig:Comp}, and the dark-dim state connection we have depicted in Fig. \ref{fig:Q} (c). A faint peak is just about visible in the third quarter of the phase space, establishing the radius of the spiral span due to the external drive. It is remarkable that all this information can be extracted from the {\it instantaneous} \textit{quasi}-distribution function for the intracavity field alone, after the qubit degrees of freedom have been traced out.

\begin{figure*}
\centering
\includegraphics[width=6.5in]{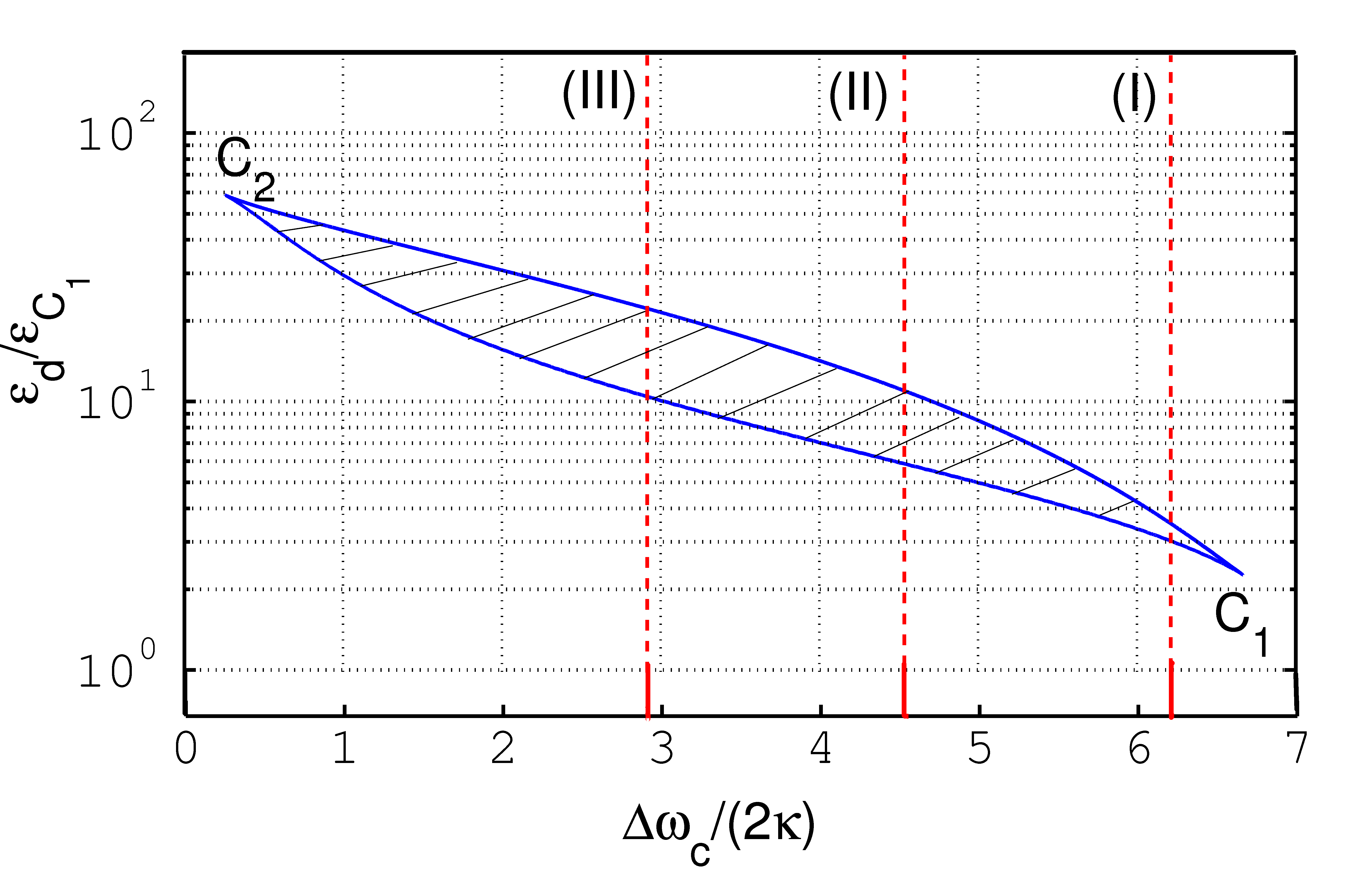}
\includegraphics[width=6.0in]{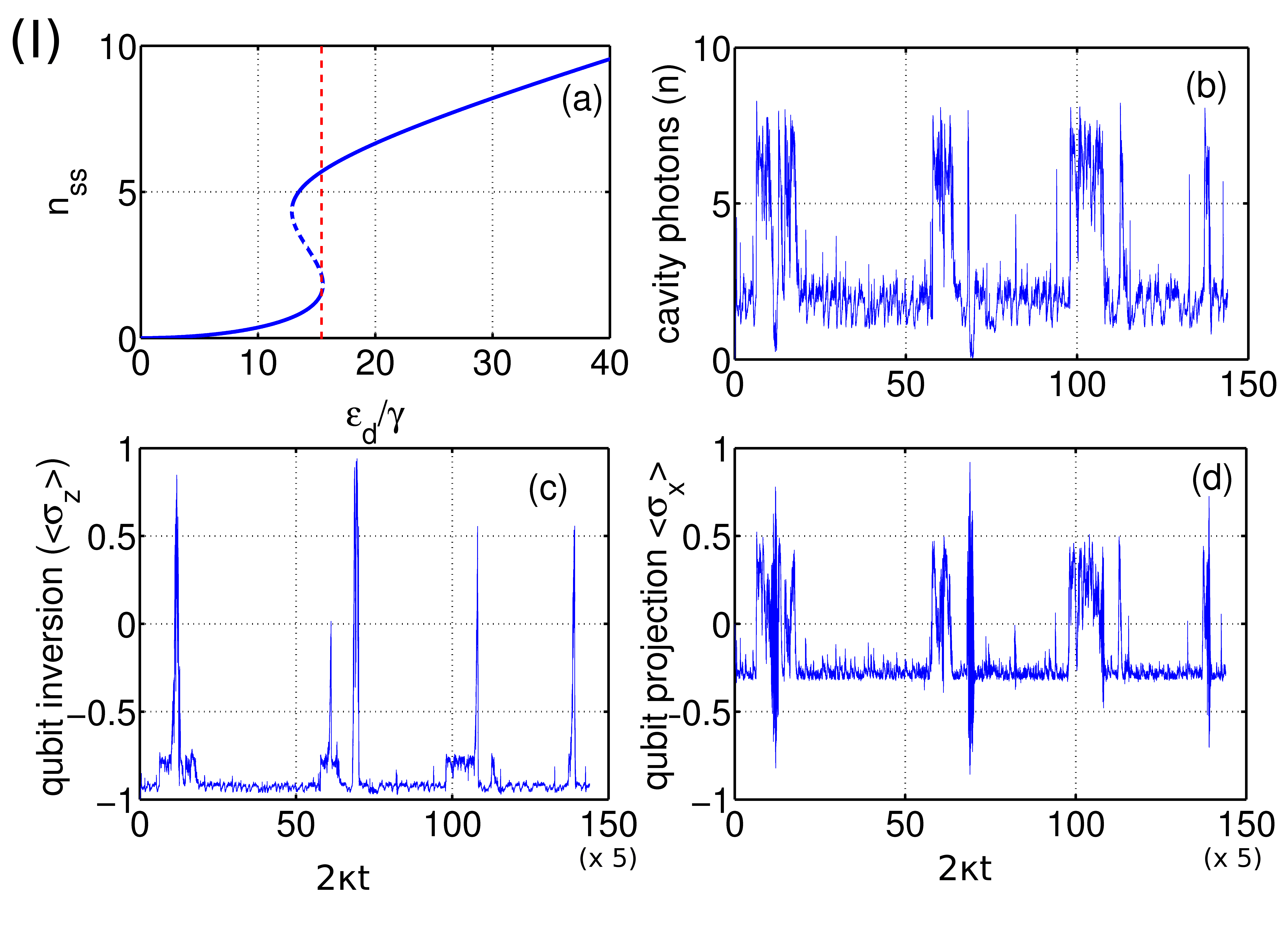}
\end{figure*}

\begin{figure*}
\includegraphics[width=6in]{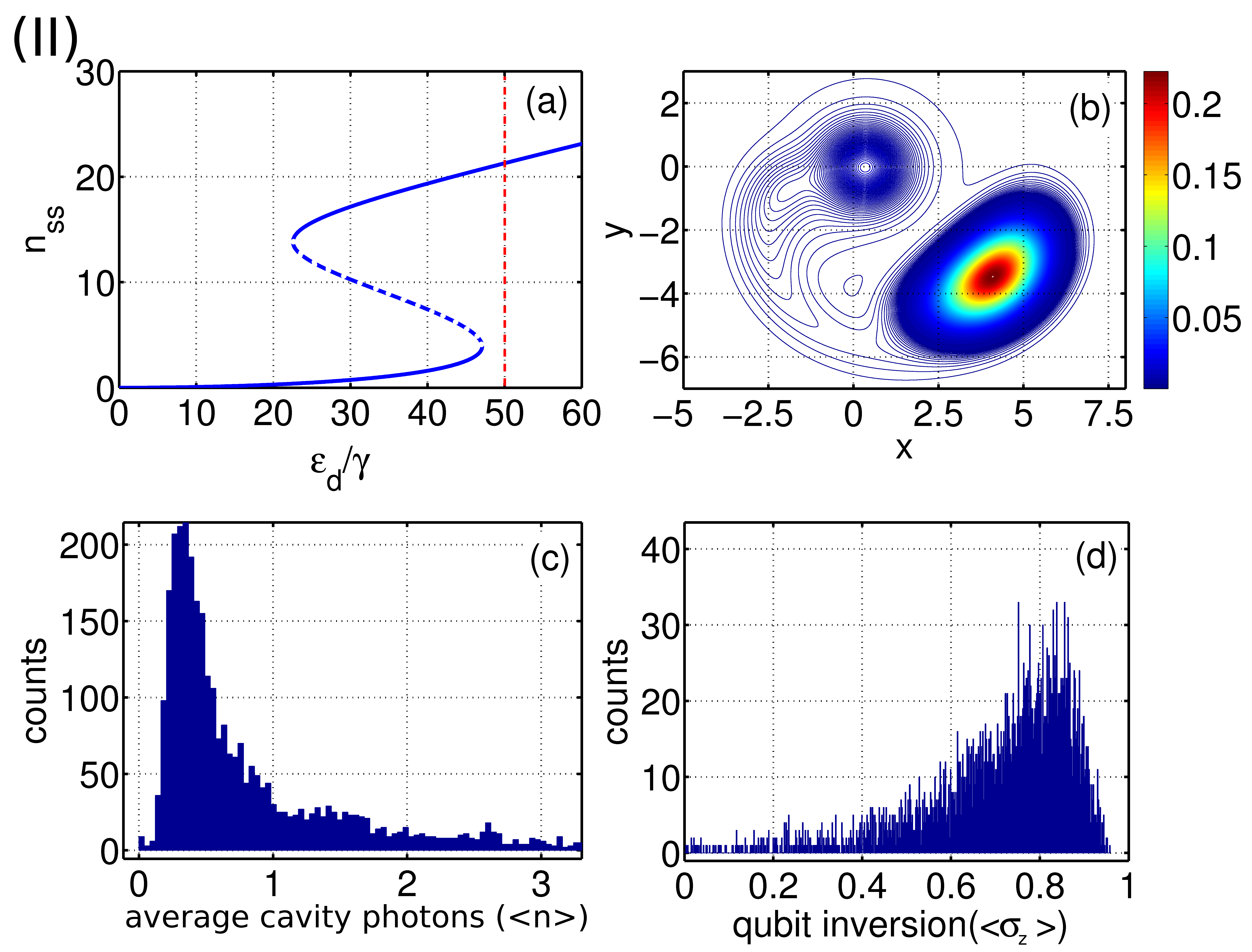}
\includegraphics[width=6in]{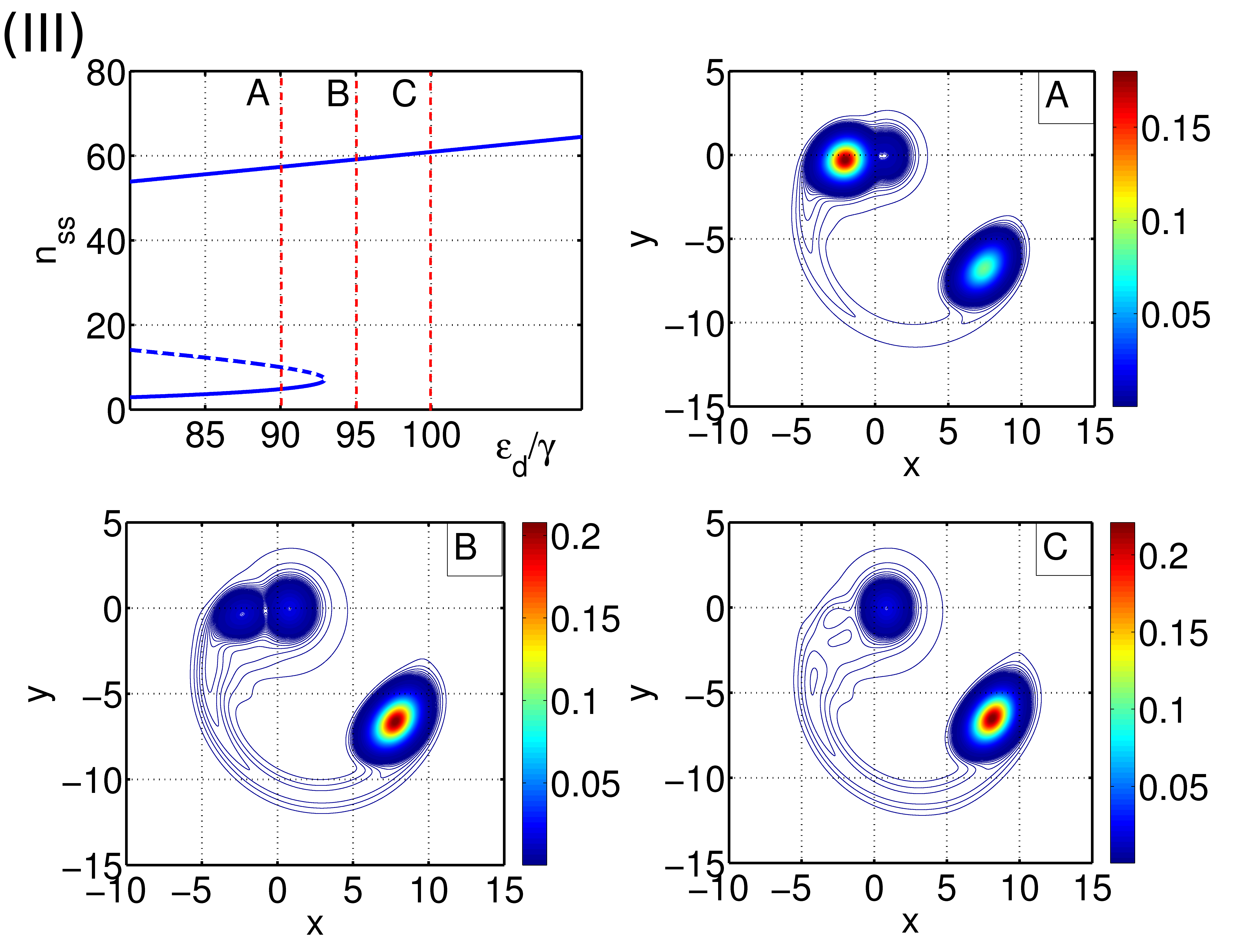}
\end{figure*}

\begin{figure}
\caption{\textbf{Drive parameter phase diagram and the dispersive JC bistability.} Bistability leaf produced from Hamilton's equations, assuming $\sigma_z$ is a constant of motion (see Fig. 2(a) of \cite{BishopJC}). The two critical points $C_{1,2}$ are marked together with three vertical cuts (I-III) following the transition from $C_1$ to $C_2$. Each of the three cuts corresponds to a panel given underneath for a different value of the cavity detuning. \underline{Panel I}: \textbf{(a)} Low-power Maxwell-Bloch bistability curve with the drive strength marked by the red dashed line. \textbf{(b, c, d)} Single quantum trajectory depicting the intracavity photons (a), the qubit inversion $\braket{\sigma_z}$ (b) and the qubit projection $\braket{\sigma_x}$ as a function of the dimensionless time $2\kappa t$ (measured against the cavity linewidth). \underline{Panel II}: \textbf{(a)} Middle-power Maxwell-Bloch bistability curve with the drive strength marked by the red dashed line. \textbf{(b)} Contour plot of the {\it quasi}-distribution function $Q(x + iy)$ for the intracavity amplitude. \textbf{(c, d)} Histogram depicting the statistical distribution of the dark state for the intracavity photons (c) and the qubit inversion $\braket{\sigma_z}$ calculated from a single quantum trajectory. \underline{Panel III}: High-power Maxwell-Bloch bistability curve with three increasing values of the drive strength marked the red dashed lines and the letters (A, B, C). For each of the drives A, B, and C, we give contour plots of the {\it quasi}-distribution functions $Q(x + iy)$ for the intracavity amplitude obtained from the solution of the ME, designated accordingly. The dashed part of the mean-field curves marks the unstable branch. For the underlying steady-state equations see Section 11.1 of \cite{WallsBook}. Parameters: $g/\gamma=600$, $2\kappa/\gamma=12$.}
\label{fig:3cuts}
\end{figure}
The cavity nonlinearity manifests in a non-perturbative fashion only in the region of large photon numbers in comparison to $n_{\rm scale}$, far within the semiclassical bistability region. In Fig. \ref{fig:ME2f} we show the variation of the moduli of the complex cavity amplitude and the qubit projection $\left|\braket{\sigma_{-}}\right|$ as a function of the normalized drive phase space. The deformation of the Lorentzian shape is accompanied by a line in the phase space where the complex amplitudes of the two metastable states cancel coherently, as we can see in Fig. \ref{fig:ME2f}. In that region of pronounced quantum fluctuations, the correlation function $g_{\rm ss}^{(2)}(\tau=0)$ attains its maximum, a behavior which is also a discerning feature of the Duffing oscillator \cite{DuffingWalls}.

A limiting behavior, in the sense discussed in \cite{PhotonBlockade}, is achieved for $g \to 0$ (sending $n_{\rm scale}$ to infinity). This is a weak-coupling limit [with the co-operativity parameter $C=g^2/(\kappa\gamma)$ remaining constant] for which the fluctuations vanish and the ME results are in close agreement with the mean-field predictions. Nonlinearity manifests itself markedly differently, however, in the case of drive-cavity resonance, where $\omega_c=\omega_d$: along the line $\Delta\omega_c=0$ the semiclassical amplitude bistability region closes up in contrast to the Duffing oscillator phase diagram \cite{BishopJC}. The emerging phase bistability for $\delta=0$ is associated with spontaneous symmetry breaking alongside a second order phase transition, more evidence of the JC nonlinearity \cite{PhotonBlockade, CarmichaelBook2}. Amplitude bistability is again recovered in the presence of spontaneous emission for $\gamma \gg 2\kappa$, with a sufficiently large saturation parameter $\gamma^2/(8g^2)$, as shown in \cite{Savage}. In our case, however, this number is vanishingly small, and $n_{\rm scale}$ sets the dominant scale for the manifestation of dispersive nonlinearity. 
\begin{figure}
\centering
\includegraphics[width=3.7in]{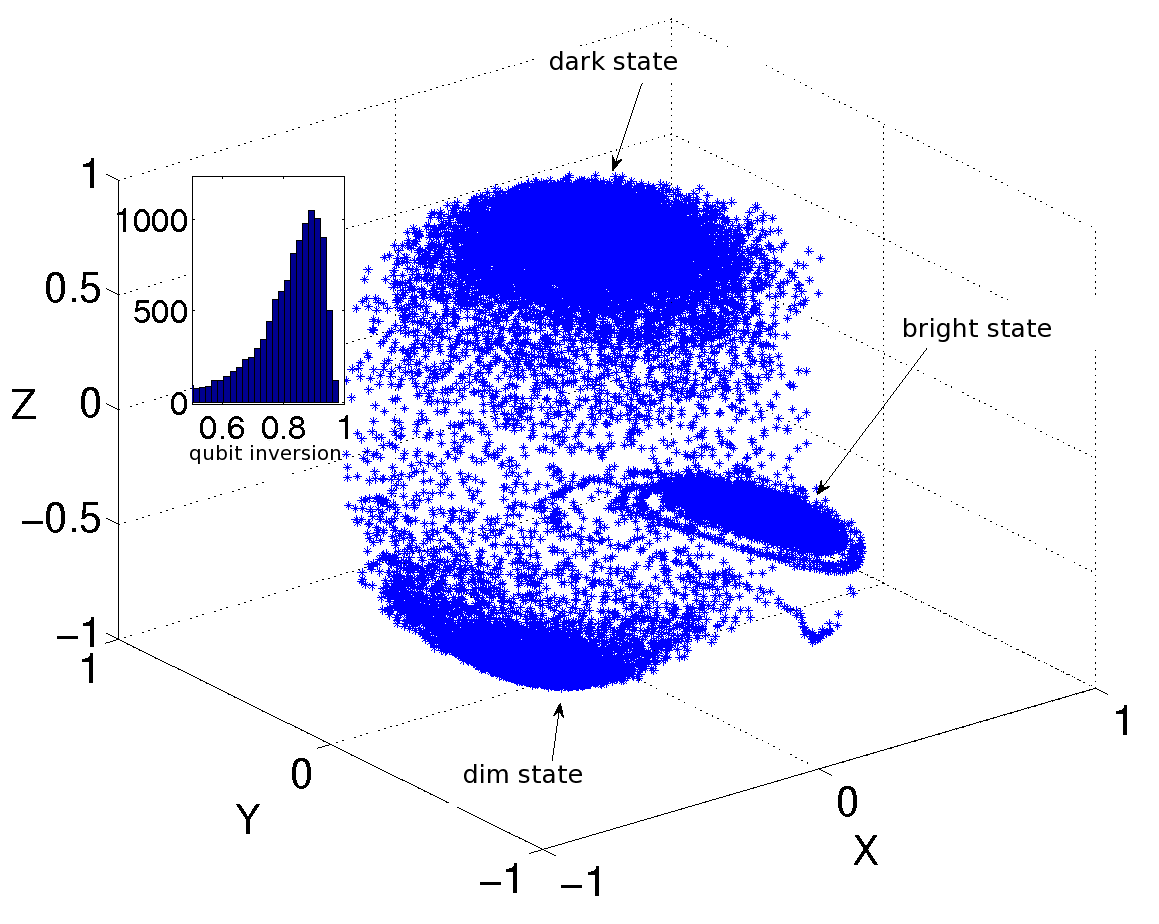}
\caption{\textbf{Qubit switching dynamics with $\gamma \neq 0$.} Bloch sphere scatter-plot from a single quantum trajectory, with each point corresponding to a particular time instant. The inset depicts the histogram of $\Braket{\sigma_z}$ generated from the time period when the system is in the dark state. Parameters: $\varepsilon_d/\kappa=16.67$, $g/\delta=0.14$, $\gamma/(2\kappa)=1/12$, $g/\gamma=3347$.}
\label{fig:SupF2A} 
\end{figure}
The mean-field states of dispersive bistability are sensitive to both the initial conditions and the value of $\gamma$. Different phase-space portraits for the intracavity field are depicted in Fig. \ref{fig:SupF3} for varying $\gamma$ and initial conditions. In Fig. \ref{fig:SupF3}(a), the initial conditions have been selected to be very close to the dim steady-state amplitudes. For $\gamma \neq 0$ and $\gamma=0$ we see an approach to the dim-state fixed point, which is however lost if the initial conditions change. A new state is approached in Fig. \ref{fig:SupF3} (c) for $\gamma=0$, different from the metastable states of dispersive bistability, which are recovered for $\gamma \neq 0$. 
\begin{figure}
\centering
\includegraphics[width=3.7in]{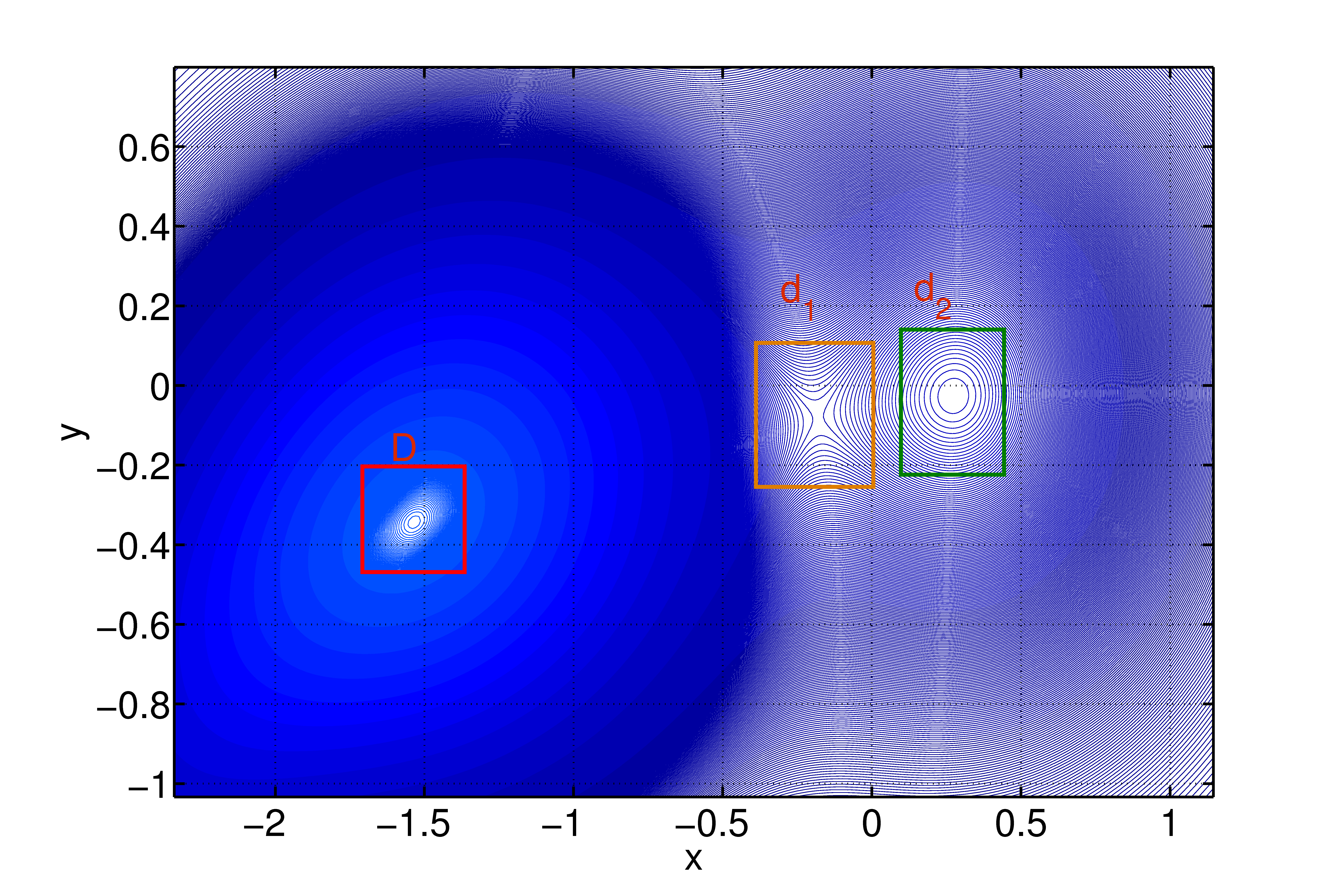}
\caption{Detail of the Wigner function contour plot $W(x + iy)$ for $\varepsilon_d/\gamma=45$ corresponding to Fig. \ref{fig:WignerBist}(d). The dim state is denoted by $D$, the unstable node by $d_1$ and the center by $d_2$. Parameters: $\Delta\omega_c/\kappa=9.167$, $g/\gamma=600$, $2\kappa/\gamma=12$.}
\label{fig:SupF5} 
\end{figure}
We are guided by the mean-field results to explore the attributes of the dark state for three different drive strengths using single quantum trajectories and the (averaged) exact ME results, depicted in Fig. \ref{fig:3cuts} (Panels I-III). The frames (c) and (d) attest that the dark state is characterized by intense qubit fluctuations which follow {\it quasi}-Poissonian statistics, to which the frame (d) of Panel II testifies. At the same time, there is clear evidence of the dark state from the ME steady-state cavity distribution in frame (b) of Panel II showing a particular excitation path linking the dark to the bright state, which is the only one anticipated by the Maxwell-Bloch equations. The dark state coexists with the dim state in frames A and B of Panel III in Fig. \ref{fig:3cuts} until the dim state vanishes completely into the excitation probability path for increased drive strength. 
\begin{figure*}
\centering
\includegraphics[width=6.5in]{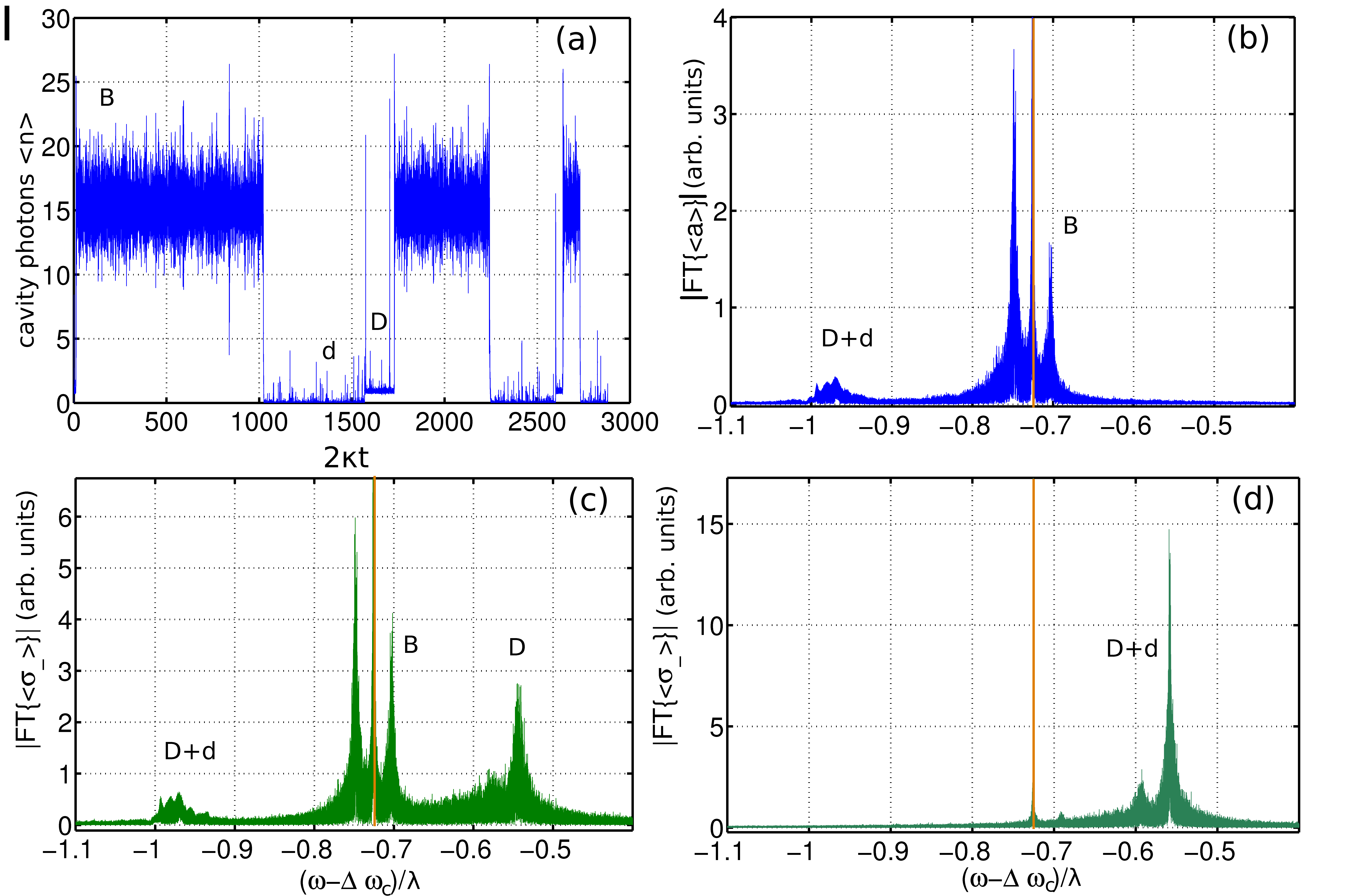}
\includegraphics[width=6.5in]{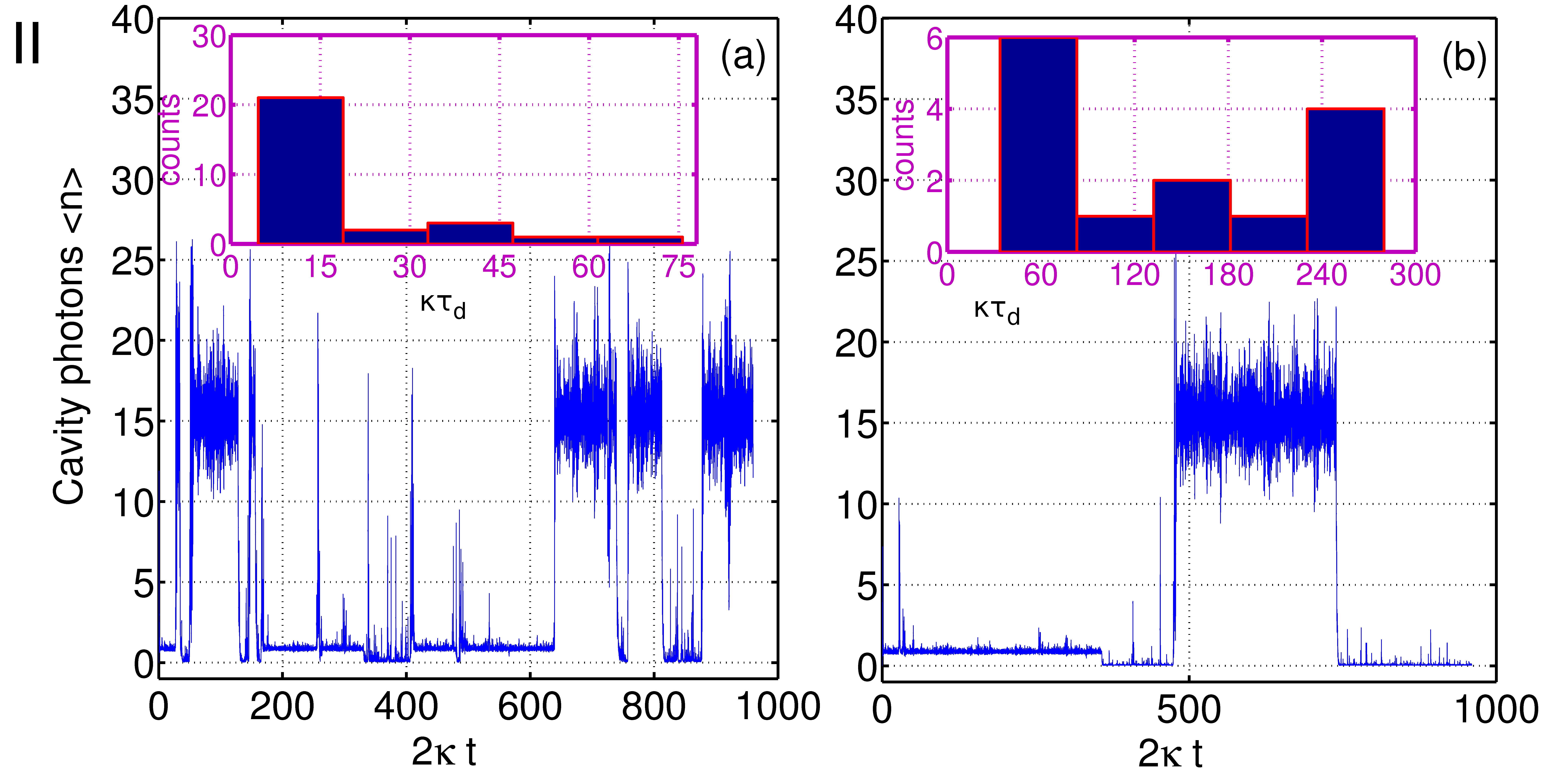}
\caption{\textbf{Spectrum and lifetime of the dark state.} \underline{Panel I}: \textbf{(a)} Photon
  number for a single quantum trajectory. \textbf{(b)} The
  corresponding (magnitude of the) Fourier Transform of the coherent intracavity field
  $\braket{a(t)}$ in the steady state. \textbf{(c)} The corresponding (magnitude of the) Fourier Transform of the coherent qubit average
  $\braket{\sigma_{-}(t)}$ in the steady state. \textbf{(d)} Magnitude of the Fourier Transform of $\braket{\sigma_{-}(t)}$ for a lower drive, giving rise
  to switching between the dim and dark states only. The orange line
  marks the frequency of the drive. Here, $\lambda=g^2/\delta$ is the
  dispersive shift. Parameters: $\gamma=0$, $g/(2\kappa)=279$,
  ${\varepsilon_d}/(2\kappa)=100/12$ for (a-c) and
  ${\varepsilon_d}^{\prime}/(2\kappa)=77/12$ for (d). \underline{Panel II}: Sample quantum trajectory and (dimensionless) lifetime ($\kappa \tau_d$) histogram of the dark state for $\gamma/(2\kappa)=0.21$ in \textbf{(a)} and $\gamma/(2\kappa)=0$ in \textbf{(b)}. In (a) we can find more frequent yet short-lived occurrences of the dark state. In (b) the effective lifetime of the qubit is limited by the Purcell decay. Parameters: $g/(2\kappa)=279$, ${\varepsilon_d}/(2\kappa)=100/12$ and $g/\delta=0.14$.}
\label{fig:FFTs} 
\end{figure*}
A quick look at the Bloch sphere of Fig. \ref{fig:SupF2A} suffices to convince us of the departure from the mean-field predictions. According to the Maxwell-Bloch equations, we would expect to find the qubit vector lying solely on the southern hemisphere of the Bloch sphere. Interestingly, these fluctuations are described by {\it quasi}-Poissonian statistics as well, with a mean inversion in the northern hemisphere.

To conclude this section, we will present the dim state and the two nodes identified as the dark state away from $C_1$, in connection to Fig. \ref{fig:WignerBist}. Figure \ref{fig:SupF5} shows the Wigner function in a drive region where the qubit participates significantly in the dynamics, exhibiting large fluctuations in the Bloch sphere when transitioning between the dim and the dark states. 

In the following section we will investigate the mean-field dispersive bistability in the absence of spontaneous emission, prompted by the behavior we have encountered in Figs. \ref{fig:Comp} and \ref{fig:SupF5}. A change is heralded by a new scaling parameter relevant for the development of nonlinearity, namely $\delta^2/(4g^2)$, as the equation for dispersive bistability shows:
\begin{equation}\label{KerrDisp}
\alpha=-i \varepsilon_d \left\{\kappa - i\left[\Delta \omega_c - \frac{g^2}{\delta} \left(1 + \frac{4g^2}{\delta^2}|\alpha|^2\right)^{-1/2}\right] \right\}^{-1}
\end{equation}
when $\gamma=0$ \cite{PhotonBlockade}. 

\section{Switching dynamics in single quantum trajectories}
\label{sec:switchdyn}

Let us now seek some evidence of the dark state within the bistable
switching itself, as a result of the quantum fluctuations. We have
performed the ME unraveling through numerically solving Stochastic
Schr\"{o}dinger Equations (SSEs) using the second-order weak scheme in
the diffusive approximation, as devised by Platen (see Ch. 15 of
\cite{PlatenBook}). The presence of the dark state is associated with
an intense fluctuation having a spectral content on the left of the
drive frequency, far beyond the spectral peaks of the bright and the
dim states. The spectrum of the coherent fields $\braket{a(t)}$ and
$\braket{\sigma_{-}(t)}$ is expected to be asymmetric with respect to
the drive, because of the presence of dissipation. The dark state
reinforces this asymmetry. When $\gamma=0$ the lifetime of the dark
state is significantly prolonged and comparable to that of the
metastable states [Fig. \ref{fig:FFTs} (a)]. As the spectra of frames
(b,c) in Fig. \ref{fig:FFTs} evidence, when the quantum fluctuation
switching involves the dark state, there is a peak in the spectrum located at
$-g^2/\delta$ in the rotating frame (excluding the time evolution independent of cavity-qubit coupling),
corresponding to the qubit flipping from a state with
$\braket{\sigma_z}=-1$ to a state with $\braket{\sigma_z}=+1$. At the
same time, the appearance of the bright state is a key element to the
switching. In Fig. \ref{fig:FFTs}(d) we encounter a
situation where only the dark and the dim states are present, in which
case there are no peaks on the left of the drive tone.
\begin{figure}
\centering
\includegraphics[width=3.6in]{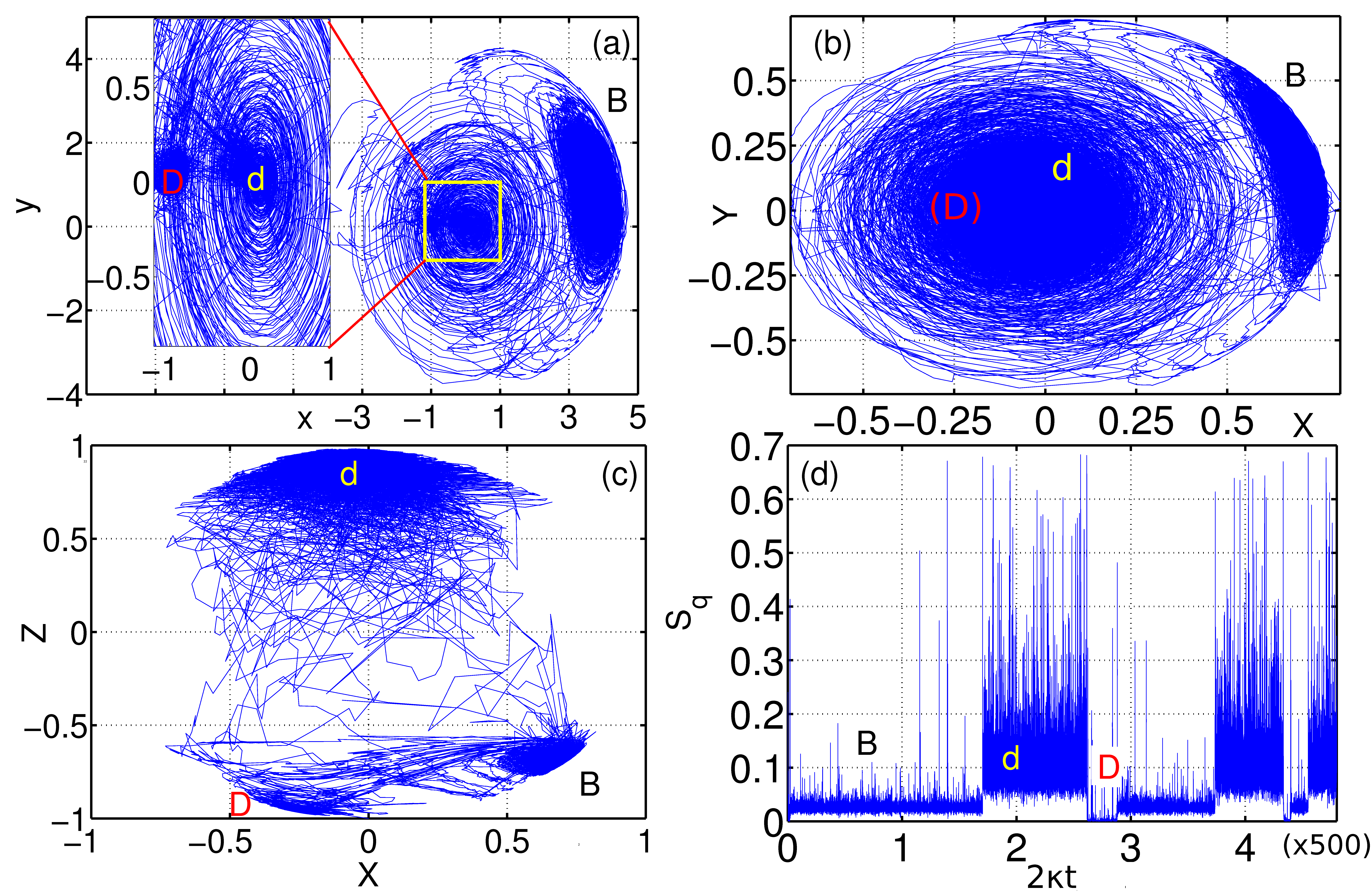}
\caption{\textbf{Bistability and entanglement in a single quantum
    trajectory for $\gamma=0$.} \textbf{(a)} Phase portrait (with $\alpha=x+iy$) of the cavity field $\braket{a^{\dagger}(t)}$ . \textbf{(b)} Qubit trajectory in the $X-Y$ plane of the Bloch sphere (depicting $\braket{\sigma_{+}(t)}$). \textbf{(c)} Qubit trajectory in the $X-Z$ plane of the Bloch sphere. \textbf{(d)} The von Neumann entanglement entropy $S_q$. The letters {\bf (B, D, d)} denote the (bright, dim, dark) states, respectively. Parameters: $\varepsilon_d/\kappa=100/6$, $g/\delta=0.14$ and $g/\gamma=3347$.}
\label{fig:QT} 
\end{figure}
The qubit flipping brings the cavity mode out of resonance, far
detuned from the drive frequency. In the low-amplitude dispersive
regime, the cavity response is a Lorentzian centered at
$\Delta\omega_c= g^2/\delta$. Hence, for $\Delta \omega_c>0$ and
higher drive strengths, the appearance of the dark state can be
construed as a spontaneous projection to a state with
$\braket{\sigma_z} \approx +1$ and very low intracavity excitation,
which is not an expected mean-field solution for $\gamma \neq 0$. As
the dark state is visited (with one center and one unstable node, as
shown in Figs. \ref{fig:SupF5} and \ref{fig:positions}), the qubit inversion
transitions from $\braket{\sigma_z} \approx +1$ to $\braket{\sigma_z}
\approx -1$, which explains the observed intense fluctuations seen in the quantum trajectories. The significance of the low-power regime is further ascertained by the analytical expression for the Wigner
function we encountered for the Duffing oscillator with one
`active' quantum degree of freedom, which captures a variety of nodes apart from the Maxwell-Bloch states (see Fig. \ref{fig:Comp} in the Appendix). Switching can occur between the dim and dark states only [with the spectrum of Frame \ref{fig:FFTs} (d) in Panel I], whereas with increasing $\gamma$ the dim state dominates and the dark state appears short-lived after the bright metastable state, as depicted in Panel II of Fig. \ref{fig:FFTs}. The histograms of that panel provide information on the lifetime of the dark state, which changes from about $20$ to about $150$ cavity lifetimes, on average, with diminishing spontaneous emission rate.
\begin{figure}
\centering
\includegraphics[width=3.7in]{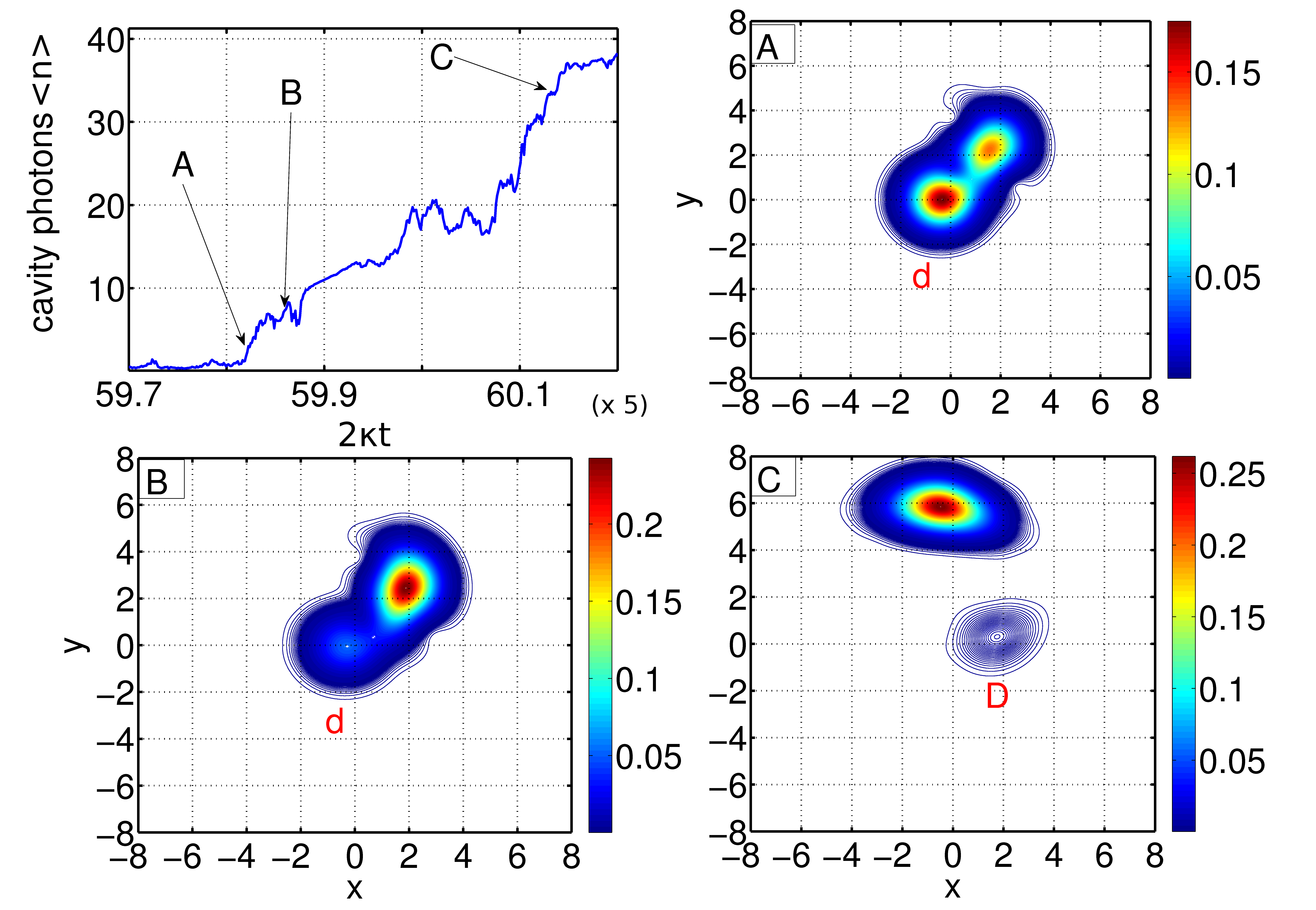}
\caption{\textbf{Switching from the dark to the bright state.} Single quantum trajectory depicting the switch from the dark to the bright state sampled at three particular time instants $t_A, t_B, t_C$ marked by the letters A-C. For each of the drives A, B, C we give contour plots of {\it quasi}-distribution functions $Q(x + iy)$ for the intracavity amplitude obtained from the numerical solution of the SSE, designated accordingly. The dark (dim) state is marked by $d (D)$. Parameters: $\delta/g=0.873$, $\Delta\omega_c/\kappa=9.167$, $g/\gamma=600$ and $\varepsilon_d/\gamma=50$.}
\label{fig:DarkTraj} 
\end{figure}
We will now focus on the phase-space representation of the dark state
during a transition involving a metastable state of the Maxwell-Bloch
bistability. Regarding the salient features of the cavity amplitude
{\it quasi}-distribution, we are already familiar with the spiral
rotation in the phase space following the de-excitation path in the JC
ladder at resonance in the presence of dissipation [see Fig. 3(b) of
\cite{PhotonBlockade} for resonance]. We are also acquainted with
squeezing in the quadrature along the mean-field direction, from
resonance fluorescence \cite{WallsZoller, CarmichaelBook1} as well as
from the Duffing oscillator \cite{CarmichaelBook2}. Switching among
metastable states means another swirl in the spiral established by
intracavity ($g$) and intercavity ($\varepsilon_d, \kappa, \gamma$) coupling,
combining features of resonance fluorescence and decaying optical
oscillations. On the one hand, such a representation reveals the
statistical mixture of semi-coherent states involved in the switching
itself, and on the other hand it provides details on the excitation
path followed in the JC ladder. At a particular time instance during the decay of
the unstable state to the bright state, probability accrues at the
highly-excited cavity state while the bottom part of the spiral
becomes more pronounced (in accordance with the fully-averaged results
of Fig. \ref{fig:WignerBist}). At that time we expect to find the
qubit amplitude following a trajectory that encircles the bright state
(with $\braket{\sigma_{z}}$ closer to zero than in the region of the
low-excitation critical point) in a limit cycle fashion, as shown in Figs. \ref{fig:QT} and \ref{fig:SupF2A}. While this spiral is described,
the dark state is occasionally visited, when the qubit vector is found
on the north pole of the Bloch sphere.
\begin{figure}
\centering
\includegraphics[width=3.5in]{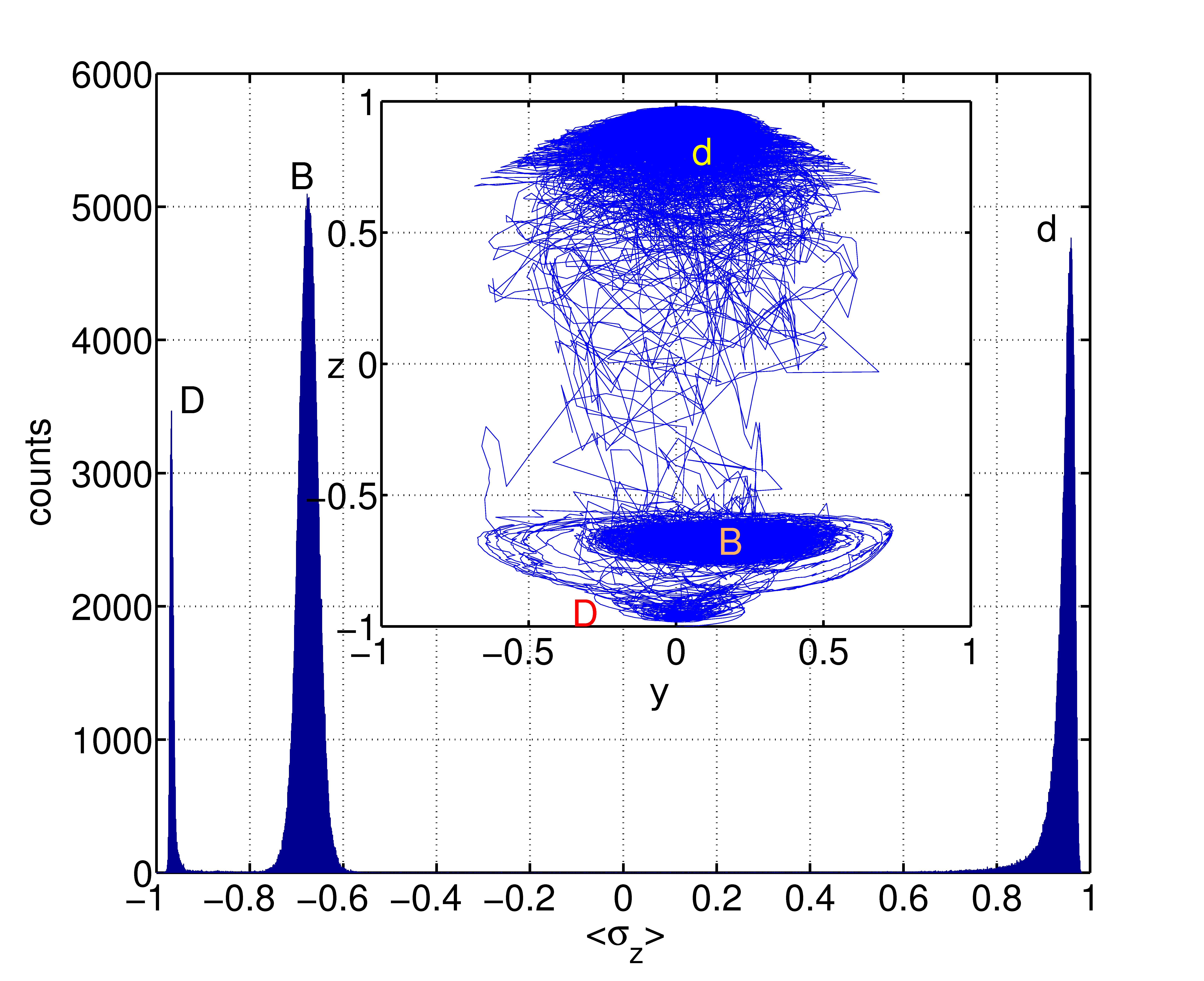}
\caption{\textbf{Qubit switching dynamics with $\gamma=0$.} Dark state histogram alongside the dim and the bright metastable states. Inset: Qubit trajectory in the $y-z$ plane of the Bloch sphere. The letters {\bf (B, D, d)} denote the (bright, dim, dark) states, respectively. Parameters: $\varepsilon_d/\kappa=100/6$, $g/\delta=0.14$, $\gamma/(2\kappa)=0$, $g/\gamma_P \sim 10^4$ ($\gamma_P$ is the Purcell decay rate).}
\label{fig:DSpin}
\end{figure}
In Figure \ref{fig:QT} we show joint bistability for the cavity and
qubit, when the bright and dim state distributions are significantly
separated. The entanglement entropy attains its highest values during
the occupation of the dark state [see Fig. \ref{fig:QT}(d)], which is
consistent with the breakdown of the Duffing approximation and the
description provided by the Maxwell-Bloch equations, due to the active
participation of both quantum degrees of freedom. In most cases, the
dark state follows the transition from the bright to the dim
metastable state, with a lifetime which is much shorter than the
duration of the two metastable states in the presence of spontaneous
emission, yet significant in comparison to the cavity and qubit
lifetimes, as Fig. \ref{fig:QT} evidences (for a more detailed account
on the switching, see Secs. \ref{subsec:DarkstateIncD} and \ref{subsec:neocleq}).
\begin{figure}
\centering
\includegraphics[width=3.0in]{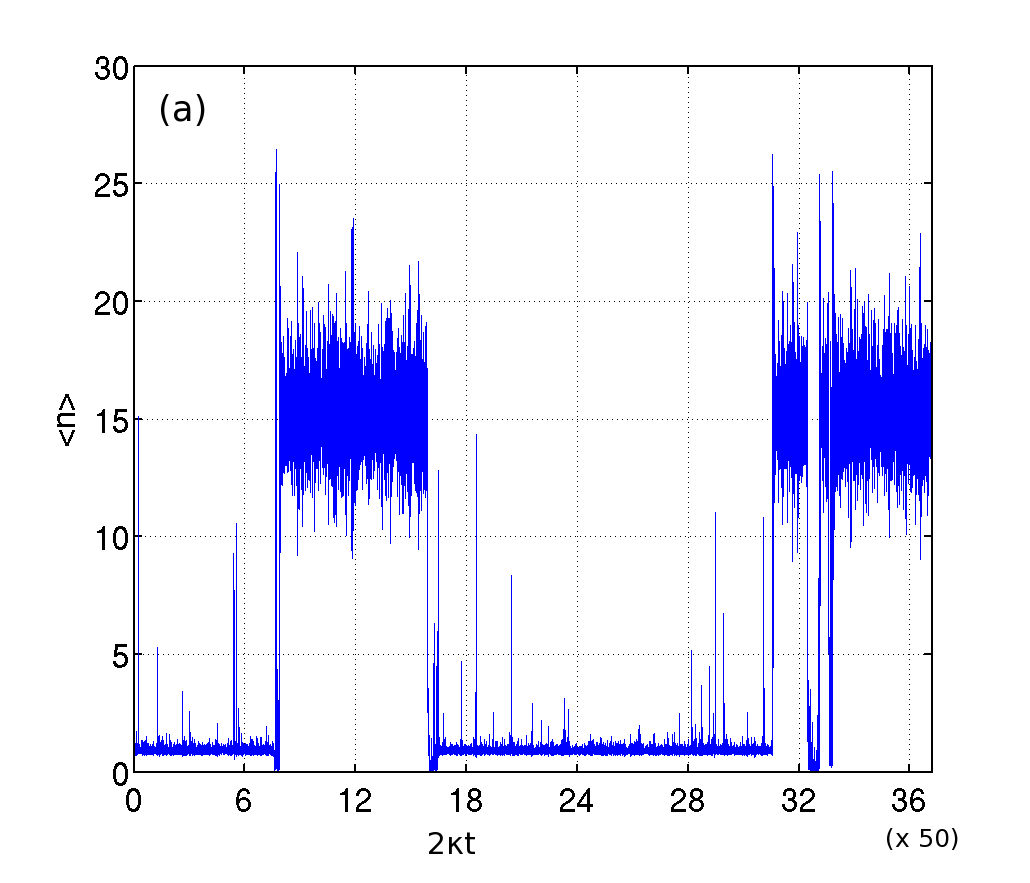}
\includegraphics[width=3.0in]{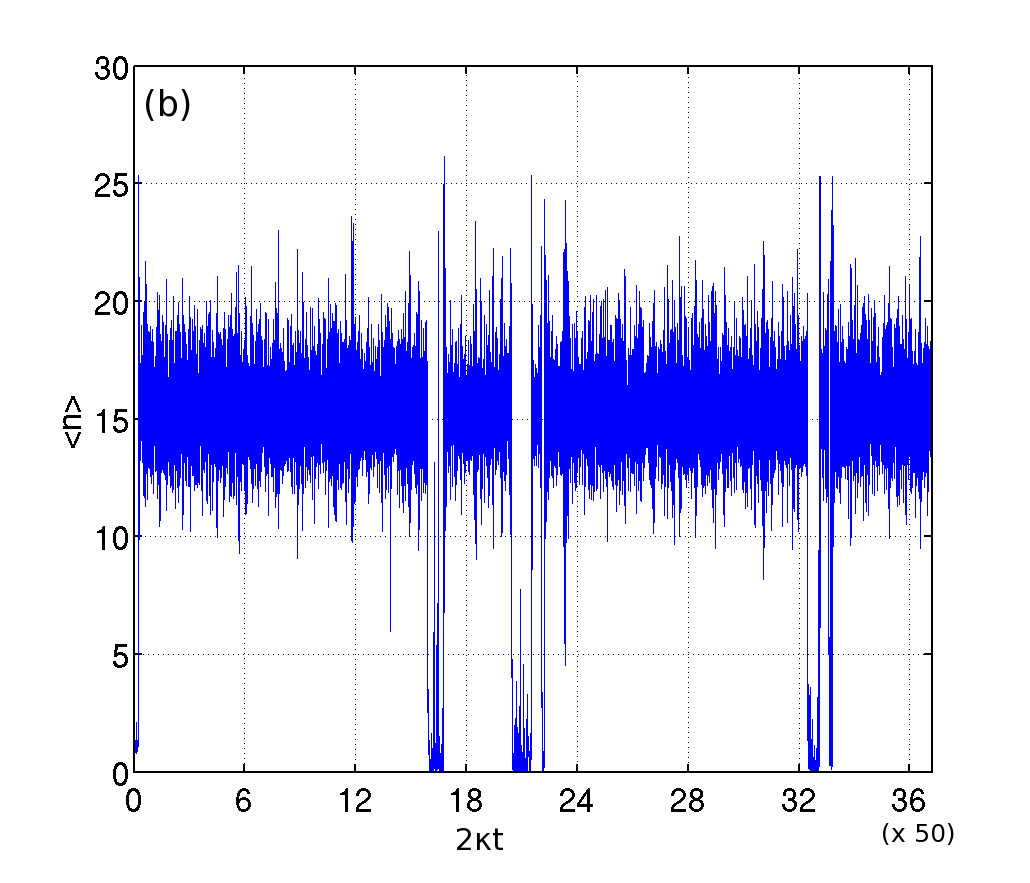}
\caption{Cavity photons in a single quantum trajectory for $\varepsilon_d/\gamma=100, 104$ in (a), (b) respectively. Parameters: $\Delta\omega_c/\kappa=56.833$, $g/\gamma=3347$, $2\kappa/\gamma=12$.}
\label{fig:SupF6} 
\end{figure}
In Fig. \ref{fig:DarkTraj}, we are following the thread from the
appearance of the dark state up to the establishment of the bright
metastable state, in a regime where the Maxwell-Bloch equations
predict only the occurrence of the latter (see Fig. \ref{fig:3cuts}). However, as the ME results suggest, the dark state complex comes about together with a center and an unstable point in
the dynamical evolution with changing relative position between them
during the quantum-activated switching (see Fig. \ref{fig:positions}). The dim state reappears in Fig. \ref{fig:DarkTraj}(d) as the bright state is about to be reached, presenting itself as a relic of Maxwell-Bloch dispersive
bistability. This finding supports the argument that the dark
state is fragile and subject to rare quantum fluctuations, coexisting
with the mean-field metastable states during the switching, and
vanishing over the much longer lifetimes of the latter. The
coexistence of the dark with the dim state is also verified by
the ME results.

In light of the lengthening of the dark state lifetime in the absence
of spontaneous emission, as shown in Fig. \ref{fig:FFTs},
the limit $\gamma/(2\kappa) \to 0$ deserves a special consideration. For
low drive strengths, the {\it quasi}-probability distribution for the
cavity field remains essentially unchanged as $\gamma \to 0$, pointing
to the fact that the nascence of complex amplitude bistability is not
related to the scale parameter $\gamma^2/(8g^2) \to 0$, as is the case
for the absorptive bistability at resonance \cite{Savage}, but rather
to $\delta^2/(4g^2)$.

A deviation from the equilibrium configuration has already been pointed out
for a multiple-atom saturable absorber on resonance, following an
adiabatic elimination of the atomic variables \cite{Bonifacio}, where
a recourse to the Gaussian probability distribution is sought for the calculation of moments, apart from the transition region where the
zero-delay second-order correlation function $g^{(2)}_{\rm ss}(\tau=0)$
diverges. In the dispersive regime, as the number of system
excitations increases, not only are we unable to assume
$\sigma_z=\braket{\sigma_z} \approx -1$, but the coupling to the
environment and drive field (and consequently the entire ME), are
rescaled to account for the actively participating system degrees of
freedom (see Sec. III of \cite{DispersiveTransform}). The dark
state appears as a result of joint quantum bistability, having roots
in the region of the critical point $C_1$ where the qubit dresses the cavity with a weak nonlinearity (see the Appendix). The lifetime of this quasi-metastable state is heavily dependent on the spontaneous emission rate, responsible for significant mixing between the various states participating in the switching. When $\gamma=0$, transitions between the qubit states occur via the cavity through the Purcell decay, with a weaker mixing, resulting in a close to a hundredfold increase in the participation of the dark state in the dynamics (for a discussion on ladder switching at resonance, see Sec. V of \cite{PhotonBlockade}). Both for zero and non-zero spontaneous emission rates, the dark state persists past the bifurcation point of semiclassical bistability (the characteristic $S$-shaped curve), where the Maxwell-Bloch equations predict the sole presence of the bright state. With increasing drive strength the dim and the dark states exchange probability and the former eventually dissolves into quantum fluctuations (see Figs. \ref{fig:WignerBist} and \ref{fig:3cuts}).

\subsection{Dark state for increasing drive strength}
\label{subsec:DarkstateIncD}

The dark state is characterized by very low photon occupation, and appears to follow closely the switching from the dim to the bright metastable state and vice versa (see for example Panel I of Fig. \ref{fig:3cuts}, as well as the Bloch-sphere phase portrait of Fig. \ref{fig:DSpin}). The associated histogram reveals a {\it quasi}-coherent state rather than a thermal state, as we can observe in Fig. \ref{fig:DSpin}. In sharp contrast with the predictions of the Maxwell-Bloch bistability, the dark state populates the upper half of the Bloch sphere.

Figure \ref{fig:SupF6} depicts a similar phenomenon to the one observed in Panel III of Fig. \ref{fig:3cuts}, where the dim state dissolves into the quantum fluctuations. In a detailed focus, the saddle point and the center of the dark state are shown in Fig. \ref{fig:positions}. 

In our treatment so far we have linked the ME results to single quantum trajectories for the cavity and qubit, showing explicitly the appearance of the dark state. This {\it quasi}-metastable state is strongly related to bistable switching between the states of mean-field dispersive bistability. This is not a necessary condition though. In many instances (see Figs. \ref{fig:3cuts} and \ref{fig:WignerBist}), we have shown bistable switching for increased drive power, where eventually the dim state disappears and only the dark state remains. In that sense, the dark state can be considered on an equal footing as the metastable states of dispersive single-atom bistability, a pure result of quantum fluctuations (see Fig. \ref{fig:FFTs} for $\gamma=0$) and remaining last in the excitation spiral after the disappearance of the dim state with increasing drive strength (see frame (d) of Panel II in Fig. \ref{fig:3cuts}). With increasing spontaneous emission rate, it comes about as a rarer fluctuation state with a shortened lifetime testifying to its fragility with respect to decoherence. At resonance, the states of phase bimodality associated with the limit $\gamma/(2\kappa) \to 0$ persist even in the presence of spontaneous emission (see Figs. 3 and 4 of \cite{SpontaneousDressedState}).

\subsection{The neoclassical equations}
\label{subsec:neocleq}

The Maxwell-Bloch equations with $\gamma=0$, also called the {\it neoclassical equations} \cite{PhotonBlockade}, predict two states lying close to the two poles of the Bloch sphere, one stable and one unstable, both having a very low photon occupation. At resonance, near the limit of zero-system size, $\gamma^2/(8g^2)=0$, the {\it neoclassical} states and the states of absorptive bistability become structurally unstable \cite{CarmichaelBook2}. A similar conclusion can be drawn for our case in the dispersive regime in the absence of spontaneous emission. We should point out here that in the dispersive regime, the above limit refers to the lower bound of the Purcell contribution (a second-order effect, see Section IV(B) of \cite{TransmonPaper} for further discussion) $\gamma_P=\kappa (g^2/\delta^2)$ which is typically one to two orders of magnitude smaller than the linear cavity decay rate in the dispersive regime. Even in the case of resonance, the limit $\gamma \to 0$ has only a formal meaning since in the absence of spontaneous emission no switching can occur between the JC excitation ladders (see Sec. 5 of \cite{SpontaneousDressedState}).

The mean-field equations of motion for $\gamma=0$, frequently called {\it neoclassical equations}, read \cite{PhotonBlockade}:
\begin{subequations}\label{neoclassical}
\begin{align}
\frac{d\alpha}{dt}&=-(\kappa-i\Delta\omega_c)\alpha - ig\mu -i\varepsilon_d, \label{eqn:fieldsc}\\
\frac{d\mu}{dt}&=i\Delta\omega_q \mu + i g \alpha \zeta, \\
\frac{d \zeta}{dt}&=2ig(\alpha^{*}\mu - \alpha \mu^{*}),
\end{align}
\end{subequations}
where $\alpha=\braket{a}$, $\mu=\braket{\sigma_{-}}$ and $\zeta=\braket{\sigma_z}=2\braket{\sigma_{+}\sigma_{-}}-1$. In the steady state, and since $\Delta\omega_q \neq 0$, we obtain
\begin{equation}\label{MF2I}
\zeta=\mp \sqrt{1 -4|\mu|^2},
\end{equation}
with
\begin{equation}\label{MF2II}
|\mu|=\frac{g|\alpha|}{\sqrt{\Delta\omega_q^2 + 4g^2 |\alpha|^2}}.
\end{equation}
\begin{figure}
\centering
\includegraphics[width=3.5in]{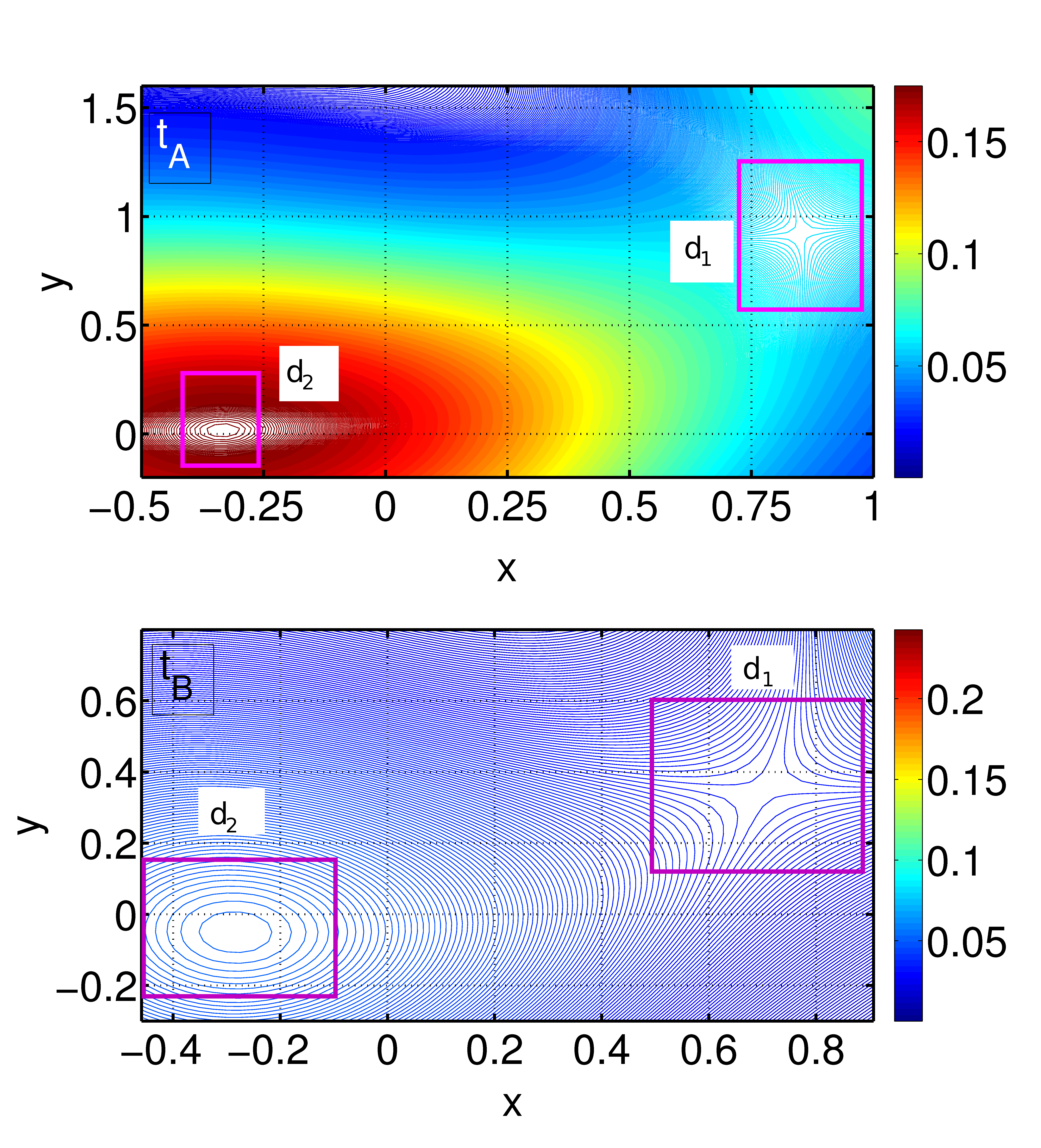}
\caption{\textbf{Pair of `fragile' nodes for the dark state.} Detail of the $Q$ function plots in the coherent state phase space for the times $t_A$ and $t_B$, given in Fig. \ref{fig:DarkTraj}. The saddle point is denoted by $d_1$ and the center by $d_2$.}
\label{fig:positions} 
\end{figure}
In the drive parameter regime under consideration, we observe that (the subscript ${\rm d}$ denotes the dark state) when $|\mu_{\rm d}|^2<(g|\alpha_{\rm d}|/|\Delta \omega_q|)^2 \ll 1$, the following inequality follows: $|\alpha_{\rm d}|^2 < [(|\varepsilon_d| + |g\mu_{\rm d}|)/|\Delta\omega_c|]^2<1$ [see Eq. \ref{eqn:fieldsc}], which suggests that the neoclassical states have an occupation below the level of one photon and also implies that one of the two neoclassical states is very close to the north pole of the Bloch sphere. As $\gamma \ll 2\kappa$ everywhere, the neoclassical state appears in the switching and coexists with the mean-field states of dispersive amplitude bistability. The Maxwell-Bloch states as well as the pair of dark states then become {\it fragile to fluctuations} as $\gamma \to 0$, in a fashion similar to the resonant case (see the relevant discussion in Sec. 16.3 of \cite{CarmichaelBook2}, and \cite{SpontaneousDressedState}). In the former reference we read, ``When $\gamma$ is close to zero, the relaxation time to these states [the steady-states of absorptive optical bistability] becomes extremely long. The limit of \textit{zero system size} is in this sense structurally unstable. It follows that near to this limit all of the mentioned states [i.e. steady-states of absorptive optical bistability and the {\it neoclassical states}] are {\it quasi-stationary} and fragile to fluctuations.'' Note also that the neoclassical solution with $\zeta>0$ is unstable with respect to fluctuations, which is yet another indication of the transient character of the dark state in our quantum simulations. The contour plots of the \textit{quasi}-distribution functions evidence the presence of two distinct states in the phase space, with very low $|\alpha|$, as predicted from Eqs. \eqref{neoclassical}. The dark state can be verified experimentally through direct Wigner tomography \cite{DirectWigner} or via observing the qubit vector close to the north pole in the Bloch sphere for a time greater than ${\rm max}\{1/(2\kappa), 1/\gamma\}$, noting at the same time the strong entanglement between the cavity and qubit.

\section{Concluding discussion}
\label{sec:conclusions}

In this work we report on the appearance of a metastable state in the
strongly dispersive regime, which is not predicted by the
Maxwell-Bloch equations. We have investigated the r\^{o}le of quantum
fluctuations, which induce bistable switching in the driven dispersive
Jaynes-Cummings model with weak spontaneous emission, in the
appropriate region of the drive strength and frequency where the
Maxwell-Bloch equations predict steady-state bistability. The
breakdown of the Duffing approximation and the appearance of terms that are 
higher order than quartic in the field operators multiplied by qubit
operators certainly suggest that an FPE cannot be formulated, with the
qubit playing a very active r\^{o}le in the cavity nonlinearity. While
some typical instances of quantum-fluctuation switching, such as the
decay of the mean-field unstable state, manifest as statistical
mixtures of semi-coherent cavity photon states with varying weights,
can we expect that the two-level atom will manifestly break the
classical picture for increasing drive powers within one quantum
trajectory? The appearance of the dark state seems to yield a
preliminary ``yes'', highlighting the importance of the neoclassical
equations combined with quantum fluctuations that are responsible for
organizing the asymptotic dynamics when $\gamma/(2\kappa)\ll 1$. The
origin of this state brings us closer to the low-excitation dispersive
regime (below the critical point $C_1$), where the qubit can be
considered a `spectator' and is not actively involved in the switching
dynamics. Fragile to fluctuations, the dark state is intimately linked
to the qubit-cavity interaction, since the entanglement entropy of the
two oscillators increases drastically while the state lasts in the
trajectory. Inasmuch as its lifetime is concerned, it may be deemed a
{\it quasi}-metastable state coexisting with the states of dispersive
bistability, which also reveals itself after the various quantum
trajectories have been averaged, in contrast to the unstable
mean-field state.

The data underlying this work is available without restriction \cite{SurreyDOI}.
 
\begin{acknowledgments} Th.~K. M.  thanks H.~J.~Carmichael for
  instructive discussions. Th.~M. and M.~H.~S. acknowledge
  support from the Engineering and Physical Sciences Research Council
  (EPSRC) under grants EP/I028900/2 and
  EP/K003623/2. E.~G. acknowledges support from the EPSRC under grant
  EP/L026082/1.
  \end{acknowledgments}

\appendix
\renewcommand\thefigure{\thesection.\arabic{figure}}    
\setcounter{figure}{0} 

\section{Wigner {\it quasi}-distribution and photon statistics in the Duffing model}

In the Appendix we derive the basic results applying to the Duffing approximation and the associated Wigner distribution we present in Section \ref{sec:DOFN}. In terms of the generalized P-representation we can write
\begin{equation}
\begin{aligned}
& W(\alpha)=\frac{2}{\pi}e^{-2|\alpha|^2} \\
& \times \int_{C_{\beta}} \int_{C_{\beta^{\dagger}}}P(\beta,{\beta}^{\dagger})\exp(2{\alpha}^*\beta + 2\alpha {\beta}^{\dagger}-2\beta{\beta}^{\dagger})d\beta d{\beta}^{\dagger}.
\end{aligned}
\end{equation}
Substituting the steady state solution for the Duffing oscillator \cite{DuffingWalls}
\begin{equation}
P(\beta, {\beta}^{\dagger})=N {\beta}^{c-2}({\beta}^{\dagger})^{d-2}\exp\left(\frac{ \tilde{\varepsilon}_d}{\beta} + \frac{{ \tilde{\varepsilon}_d}^*}{{\beta}^{\dagger}}  + 2 \beta {\beta}^{\dagger}\right),
\end{equation}
we subsequently effect the variable change (with $\tilde{\varepsilon}_d=\varepsilon_d/\chi$),
\begin{equation}
\delta= \tilde{\varepsilon}_d/\beta, \quad {\delta}^{\dagger}={\tilde{\varepsilon}_d}^{*}/{\beta}^{\dagger}.
\end{equation}
Recognizing the following integral representation of the Bessel function:
\begin{equation}
2\pi i J_{\nu}(z)=\left(\frac{z}{2}\right)^{\nu}\int_{C}t^{\nu-1}\exp\left(t- \frac{z^2}{4t}\right)\, dt,
\end{equation}
with $C$ being the Hankel path starting at $-\infty$, encircling the origin in an anticlockwise fashion, and returning back to $-\infty$, the final result reads \cite{WignerKheruntsyan, WignerKheruntsyanOneOsc}
\begin{equation}
W(\alpha)=N^{\prime} e^{-2|\alpha|^2} \left| \frac{J_{c -1}(\sqrt{-8\tilde{\varepsilon}_d {\alpha}^*})}{({\alpha}^*)^{[(c-1)/2]}} \right|^2.
\end{equation}
($N^{\prime}$ is the normalization constant). We note that the function is everywhere positive. We normalize it through the condition
\begin{equation}
\iint_{S} W(\alpha)\, d^2\alpha=1,
\end{equation}
which leads to (with $c=d^{*}$)
\begin{equation}
\begin{aligned}
&N^{\prime} \left|(-2 \tilde{\varepsilon}_d)^{c-1}\right|\displaystyle \int_0^{\infty} e^{-2\rho^2} \times \\ &\sum_{k=0}^{\infty}\sum_{l=0}^{\infty} \frac{(2{\tilde{\varepsilon}_d}{\rho})^k (2{\tilde{\varepsilon}_d}^{*}\rho)^l}{(k!) (l!) \Gamma(k+c)\Gamma(l+d)}\rho d\rho \int_0^{2\pi}e^{-i(k-l)\phi}d\phi=1.
\end{aligned}
\end{equation}
The integral over $\phi$ evaluates to $2\pi \delta_{kl}$ and the integral over $\rho$ yields
\begin{equation}
\int_0^{\infty} e^{-2\rho^2}({\rho}^2)^kd\rho=\frac{1}{2}\int_0^{\infty}e^{-2u^2}u^k\,du=\frac{k!}{2^{k+2}}.
\end{equation}
Taking into account these two results we finally arrive at 
\begin{equation}
N^{\prime}=\frac{2}{\pi \left|(-2\tilde{\varepsilon}_d)^{c-1}\right|}\frac{\Gamma(c)\Gamma(d)}{_0F_2(c,d,2|\tilde{\varepsilon}_d|^2)}.
\end{equation}
We will now prove that the probability density function $p(n)$ is normalized. From the given steady-state photon number probability density function, we have
\begin{equation}
\sum_{n=0}^{\infty}p(n)=\frac{S_0}{_0F_2(c,d,2|\tilde{\varepsilon}_d|^2)},
\end{equation}
with:
\begin{equation}
\begin{aligned}
S_0&=\sum_{u=0}^{\infty}\sum_{k=0}^{u}\frac{|\tilde{\varepsilon}_d|^{2u}}{(u-k)! k!} \frac{\Gamma (c) \Gamma (d)}{\Gamma (u+c) \Gamma(u+d)}\\
&=\sum_{u=0}^{\infty}\left(\sum_{k=0}^{u}\frac{u! (\frac{1}{2})^{u}}{(u-k)! k!}\right)\frac{(2| \tilde{\varepsilon}_d|^2)^{u}\Gamma(c) \Gamma(d)}{u! \Gamma(u+c)\Gamma(u+d)}
\end{aligned}
\end{equation}
\begin{figure}
\centering
\includegraphics[width=3.5in]{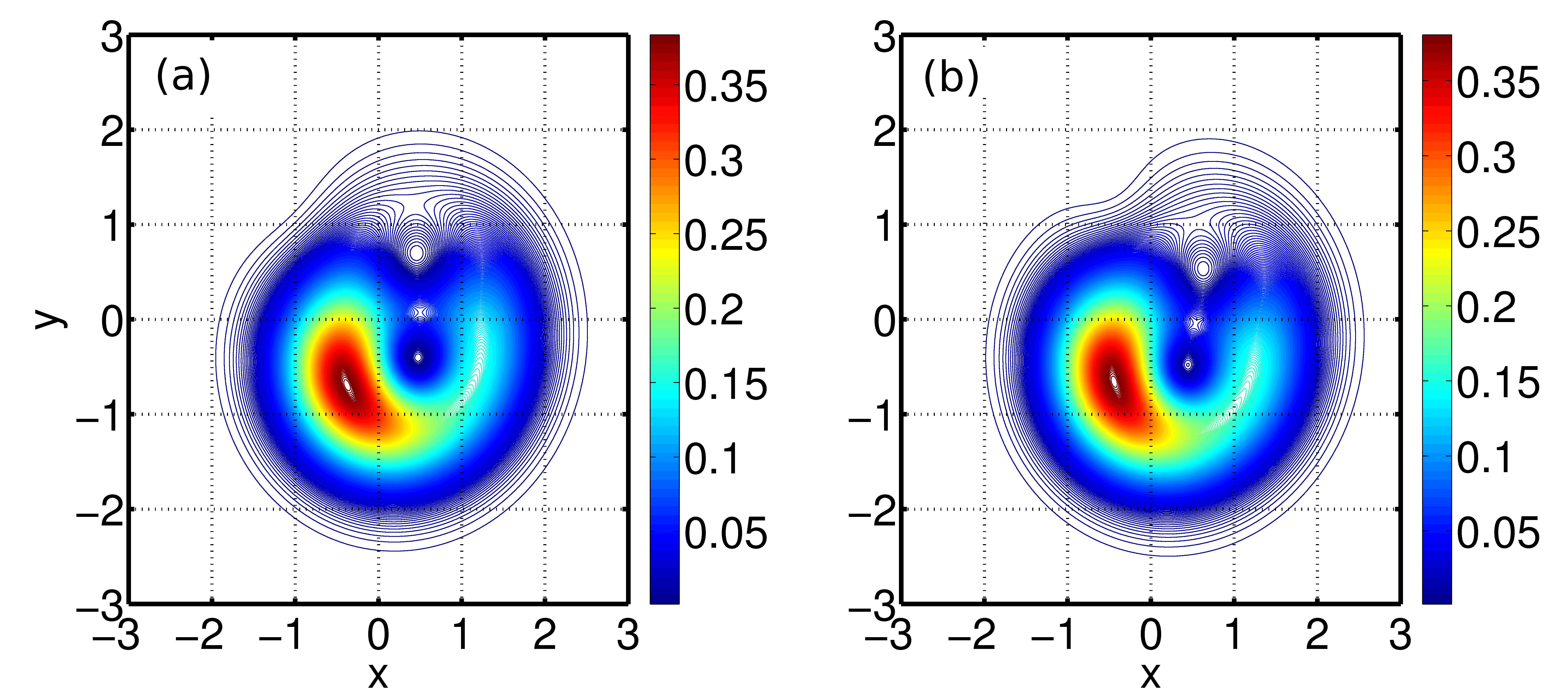}
\includegraphics[width=3.7in]{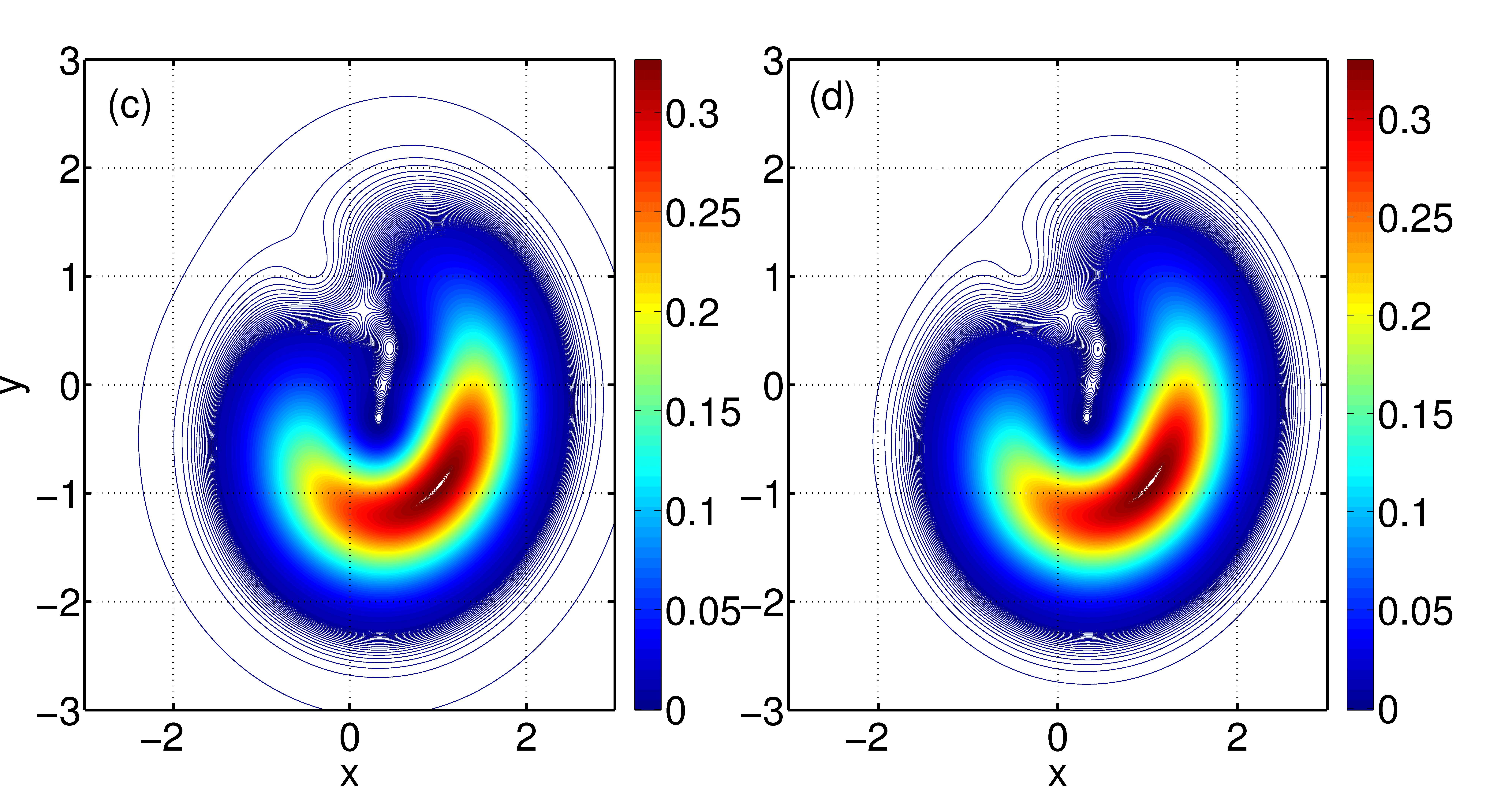}
\caption{\textbf{Effective Duffing and JC nonlinearity in the low drive strength regime.} Contour plots of the \textit{quasi}-distribution Wigner function $W(x + iy)$ for weak bistability using \textbf{(a)} the approximate Duffing reduction [Eq. \eqref{Duffingapprox}] and the JC model [Hamiltonian of Eq. \eqref{JC}] in \textbf{(b-d)}. In (b) $\gamma<2\kappa$, in (c) $\gamma=2\kappa$ and in (d) $\gamma=0$. Parameters: $\Delta\omega_c/\kappa=74.17$ $g/\delta=0.14$, $2\kappa/\gamma=12$, $g/(2\kappa) \simeq 279$, $\varepsilon_d/\kappa=1.667$ [in (a)-(b)], and $\varepsilon_d/\kappa=2.333$ [in (c)-(d)].}
\label{fig:Comp} 
\end{figure}
But the sum inside the brackets equals unity, by virtue of the normalization of the binomial distribution with $p=1/2$. Hence, the above sum reads
\begin{equation}
S=\sum_{u=0}^{\infty}\frac{(2|\tilde{\varepsilon}_d|^2)^{u}\Gamma(c) \Gamma(d)}{u! \Gamma(u+c)\Gamma(u+d)}\equiv _0F_2(c,d,2|\tilde{\varepsilon}_d|^2),
\end{equation}
which proves that $P(n)$ is normalized. For the calculation of the first moment, we have
\begin{equation}
m_1=\sum_{n=0}^{\infty}nP(n)=\frac{S_1}{_0F_2(c,d,2|\tilde{\varepsilon}_d|^2)},
\end{equation}
with
\begin{equation}
\begin{aligned}
&S_1 =|\tilde{\varepsilon}_d|^2 \sum_{u=0}^{\infty}\sum_{k=0}^{u}\frac{|\tilde{\varepsilon}_d|^{2u}}{(u-k)! k!} \frac{\Gamma (c) \Gamma (d)}{\Gamma (u+1+c) \Gamma(u+1+d)}=\\
&=|\tilde{\varepsilon}_d|^2 \sum_{u=0}^{\infty}\left(\sum_{k=0}^{u}\frac{u! (\frac{1}{2})^{u}}{(u-k)! k!}\right)\frac{(2|\tilde{\varepsilon}_d|^2)^{u}\Gamma(c+1) \Gamma(d+1)}{u! \Gamma(u+c)\Gamma(u+d)\, c d},
\end{aligned}
\end{equation}
so that finally
\begin{equation}
m_1=|\tilde{\varepsilon}_d|^2 \frac{_0F_2 (c+1, d+1, 2|\tilde{\varepsilon}_d|^2)}{cd \,_0F_2 (c, d, 2|\tilde{\varepsilon}_d|^2)}.
\end{equation}
\begin{figure}
\centering
\includegraphics[width=3.5in]{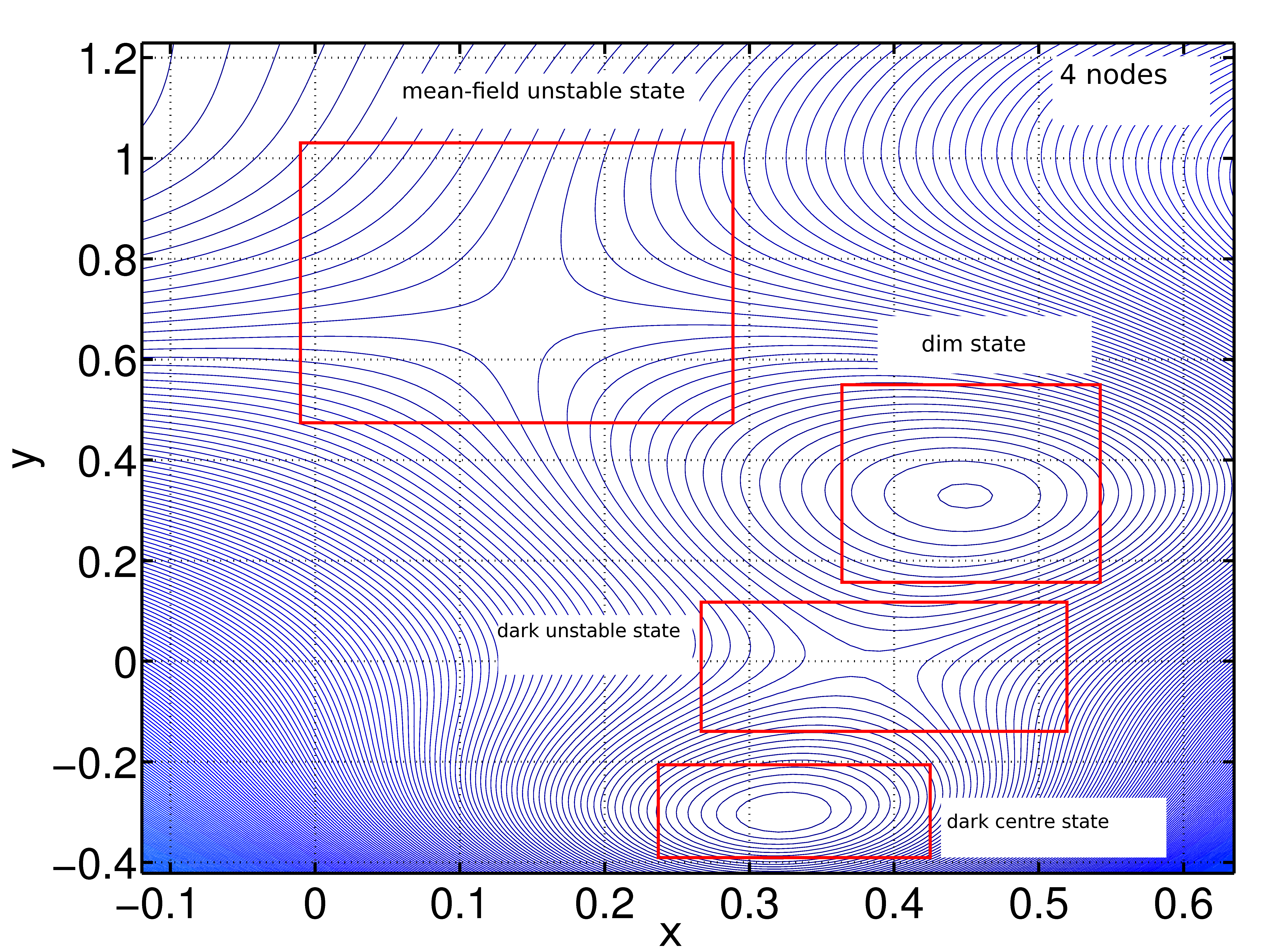}
\caption{Detail of the Wigner function of Fig. \ref{fig:Comp}(d). With red rectangles we mark two unstable and two center nodes.}
\label{fig:SupF1} 
\end{figure}
We will now use the Wigner {\it quasi}-distribution to calculate the symmetrically ordered operator moments $\Braket{({a}^{\dagger})^{n}a^{m}}_{\rm{S}}$ as follows:
\begin{equation}
\begin{aligned}
&\Braket{ ({a}^{\dagger})^{n}a^{m}}_{\rm{S}}=\frac{2}{\pi}\frac{\Gamma(c)\Gamma(d)}{_0F_2(c,d,2|\tilde{\varepsilon}_d|^2)} \sum_{k=0}^{\infty}\sum_{l=0}^{\infty}\int_0^{\infty} e^{-2\rho^2}\\ 
&\times\frac{(2{\tilde{\varepsilon}_d})^{k}{\rho}^{k+n} (2{\tilde{\varepsilon}_d}^{*})^{l}\rho^{l+m}}{(k!) (l!) \Gamma(k+c)\Gamma(l+d)}\rho\, d\rho\, \\ & 
\times\int_0^{2\pi}e^{-i(k+n-l-m)\phi}\,d\phi.
\end{aligned}
\end{equation}
The integral over $\phi$ yields $2\pi \delta_{(k+n),(l+m)}$, and thus
\begin{equation}
\begin{aligned}
&\Braket{({a}^{\dagger})^{n}a^{m}}_{\rm{S}}=\frac{2}{\pi}\frac{\Gamma(c)\Gamma(d)}{_0F_2(c,d,2|\tilde{\varepsilon}_d|^2)}\sum_{k=0}^{\infty}\int_0^{\infty} e^{-2\rho^2} \\
&\times \frac{(2{\tilde{\varepsilon}_d})^{k}{\rho}^{2(k+n)} (2{\tilde{\varepsilon}_d}^{*})^{k+n-m}}{(k!) ((k+n-m)!) \Gamma(k+c)\Gamma(k+n-m+d)}\rho d\rho.
\end{aligned}
\end{equation}
Despite the fact that the Wigner function for every steady state of the driven dissipative Duffing oscillator is positive, the departure from a Gaussian distribution is obvious in the region of bistability, which is also reflected in the expressions for the various moments of the intracavity field. The latter are highly nonlinear functions of the drive strength, as is also verified in Fig. 1 of \cite{DuffingWalls}. The corresponding steady-state photon number probability distribution function can be written as
\begin{equation}\label{eqn:Pn}
p(n)=\frac{| \tilde{\varepsilon}_d|^{2n}}{n!} \left| \frac{\Gamma(c)}{\Gamma(c+n)} \right|^2\frac{_0F_2(c+n,c^{*}+n,| \tilde{\varepsilon}_d|^2)}{_0F_2(c,c^{*},2| \tilde{\varepsilon}_d|^2)}. 
\end{equation}
The above expression shows the deviation from the Poissonian distribution of a coherent state, with increasing drive power.  

In Fig. \ref{fig:Comp}, we display the Wigner {\it quasi}-distribution functions for the intracavity photon field (with $\alpha=x+iy$) in a driving region where {\it the qubit is not significantly excited} and we can set $\sigma_z=\braket{\sigma_z}=-1$. The exact master equation predictions and the Duffing approximation of Eq. \eqref{Duffingapprox} are in good agreement, showing the development of low amplitude bistability alongside the departure from the Gaussian shape of a coherent state. Fig. \ref{fig:SupF1} details the low-amplitude region in the phase space (compare with Fig. \ref{fig:WignerBist} for a higher drive strength, far away from the point $C_1$). The cavity field {\it quasi}-distributions in Fig. \ref{fig:Comp} correspond to the photon statistics of Eq. \ref{eqn:Pn}. 

At the same time, Fig. \ref{fig:Comp}(a) reveals that the analytical expression of Eq. \eqref{Duffingapprox} already captures four nodes in the low-excitation regime: two stable and two unstable. For the drive strength used in Figs. \ref{fig:Comp}(c) and \ref{fig:Comp}(d), the mean-field analysis with $2\kappa/\gamma=12$ predicts only one state with $\braket{n}_{\rm ss} \approx 1.56$, captured in the Wigner function plots by the peak of the squeezed state centered at $\alpha_{\rm ss}\approx 1.025 - 0.92i$, exhibiting negligible variation with changing $\gamma$. The last frame of Fig. \ref{fig:WignerBist} shows that the dim state is clearly distinguished from the dark-state pair. This separation occurs at a drive strength for which the Maxwell-Bloch equations predict the existence of the bright state only, similarly to the situation we have encountered for the low-amplitude bistability approximated by the effective Duffing nonlinearity, as we have seen in Fig. \ref{fig:SupF1}.


\begin{thebibliography}{40}

\bibitem{PeanoDykman}
V. Peano and M. I. Dykman, {\it New J. Phys.} \textbf{16}, p.015011 (2014).

\bibitem{LeytonD}
V. Leyton, V. Peano, and M. Thorwart, {\it New J. Phys.} \textbf{14}, p.093024 (2012).

\bibitem{DykmanSmelyanskii}
M. I. Dykman  and V. N. Smelyanskii, {\it Sov. Phys. JETP} \textbf{67}, p.1769 (1988).

\bibitem{DykmanBook}
M. I. Dykman, {\it Fluctuating Nonlinear Oscillators}, Ch. 7, Oxford University Press (2012).

\bibitem{detailbalance}
See the conclusions and Section 5 of \cite{DuffingWalls}, as well as the subsections D(1,2) of \cite{HakenRev} for the required conditions referring to detailed balance.

\bibitem{DuffingWalls}
P. D. Drummond and D. F. Walls, {\it J. Phys. A} \textbf{13}, p.725 (1980).

\bibitem{HakenRev}
H. Haken, {\it Rev. Mod. Phys.} \textbf{47}, p.67 (1975).

\bibitem{MattFPE}
M. Elliott and E. Ginossar, {\it Phys. Rev. A} {\bf 94}, p.043840 (2016). 

\bibitem{constantD}
See \cite{Kramers} and Fig. 6 of \cite{Hanggi}.

\bibitem{Kramers}
H. A. Kramers, {\it Physica} \textbf{7}, p.284 (1940).

\bibitem{Hanggi}
P. Hanggi and P. Riseborough, {\it Am. J. Phys.} \textbf{51}, p.347 (1983).

\bibitem{CarmichaelBook1}
H. J. Carmichael, {\it Statistical Methods in Quantum Optics 1}, Springer (1999). 

\bibitem{CarmichaelBook2}
H. J.  Carmichael, {\it Statistical Methods in Quantum Optics 2}, Springer (2008). 

\bibitem{GrahamBist}
 R. Graham and  A.  Schenzle, {\it Phys. Rev. A} \textbf{23}, p.1302 (1981).

\bibitem{GrahamTel}
R. Graham  and T. T\'{e}l, {\it Phys. Rev. Lett.} \textbf{52} p.9 (1984).

\bibitem{Kamenevdis}
For a detailed discussion see Section 4.12 of \cite{Kamenev} and references
therein. 

\bibitem{Kamenev}
A. Kamenev, {\it Field Theory of Non-Equilibrium Systems}, Cambridge University Press (2011). 

\bibitem{SimBistPRL}
Th. K. Mavrogordatos,  G. Tancredi, M. Elliott, M. J. Peterer, A. Patterson, J. Rahamim, P. J. Leek, E. Ginossar, and M. H. Szyma\'{n}ska, {\it Phys. Rev. Lett.} \textbf{118}, p.040402 (2017).

\bibitem{WallsBook}
D. F. Walls and G. J. Milburn, {\it Quantum Optics}, Springer (2010).

\bibitem{PlatenBook}
P. E. Kloeden and E. Platen, {\it Numerical Solution of Stochastic Differential Equations}, Springer (1995).

\bibitem{OpenQ}
H.-P. Breuer and F. Petruccione, {\it The Theory of Open Quantum Systems}, Oxford University Press (2002). 

\bibitem{RWAval}
D. Zueco, G. M. Reuther, S. Kohler, and P. H\"{a}nggi, Phys. Rev. A {\bf 80},p. 033846 (2009).

\bibitem{RedME}
F. Beaudoin, J. M. Gambetta, and A. Blais, Phys. Rev. A {\bf 84}, p.043832 (2011).

\bibitem{BishopNL}
L. S. Bishop, J. M. Chow, Jens Koch, A. A. Houck, M. H. Devoret, E. Thuneberg, S. M. Girvin, and R. J. Schoelkopf, Nat. Phys. {\bf 5}, p.105 (2009). 

\bibitem{DispersiveTransform}
M. Boissonneault, J. M. Gambetta, and A. Blais, {\it Phys. Rev. A} \textbf{79}, p.013819 (2009).

\bibitem{CanonicalTr}
P. Carbonaro, G. Compagno, and F. Persico, {\it Phys. Lett. A} \textbf{73}, p.97 (1979). 

\bibitem{WignerKheruntsyan}
K. V. Kheruntsyan and K. G. Petrosyan, {\it Phys. Rev. A} \textbf{62}, p.015801 (2000).

\bibitem{WignerKheruntsyanOneOsc}
K. V. Kheruntsyan, {\it Journal of Optics B: Quantum and Semiclassical Optics} \textbf{1}, p.225 (1999).

\bibitem{BishopJC}
L. S. Bishop, E. Ginossar, and  S. M. Girvin, {\it Phys. Rev. Lett.} \textbf{105}, p.100505 (2010).

\bibitem{BifRoutes}
Y. Gu, D. K. Bandy, J.-M. Yuan, and L. M. Narducci, {\it Phys. Rev. A} \textbf{31}, p.354 (1985).

\bibitem{AttractorsLaser}
D. J. Jones and D. K. Bandy, {\it J. Opt. Soc. Am. B} \textbf{7}, p.2119 (1990).

\bibitem{PhotonBlockade}
 H. J. Carmichael, {\it Phys. Rev. X} \textbf{5}, p.031028 (2015).
 
\bibitem{Savage}
C. M. Savage and H. J. Carmichael, {\it IEEE J. Quant. Electron.} \textbf{24}, p.1495 (1988).

\bibitem{WallsZoller}
D. F. Walls and P. Zoller, {\it Phys. Rev. Lett.} \textbf{47}, p.709 (1981).

\bibitem{Bonifacio}
R. Bonifacio, M. Gronchi, and L. A. Lugiato, \textit{Phys. Rev. A} \textbf{18}, p. 2266 (1978).

\bibitem{SpontaneousDressedState}
P. Alsing and H. J. Carmichael, {\it Quantum Opt.} \textbf{3}, p.13 (1991).

\bibitem{TransmonPaper}
J. Koch, T. M. Yu, J. Gambetta, A. A. Houck, D. I. Schuster,
J. Majer, A. Blais, M. H. Devoret, S. M. Girvin, and R. J. Schoelkopf, {\it Phys. Rev. A} \textbf{76}, p.042319 (2007). 

\bibitem{DirectWigner}
Y. Shalibo, R. Resh, O. Fogel, D. Shwa, R. Bialczak, J. M. Martinis, and N. Katz, Phys. Rev. Lett. {\bf 110}, p.100404 (2013). 

\bibitem{SurreyDOI}
DOI: \href{http://doi.org/10.15126/surreydata.00845940}{10.15126/surreydata.00845940}

\end{thebibliography}
\end{document}